\documentclass[12pt]{article}
\pdfoutput=1
\usepackage{amsmath,amssymb,amsfonts}
\usepackage{a4wide,epsfig,psfrag,scalefnt}
\usepackage[dvipsnames]{xcolor}
\usepackage{braket}
\usepackage{placeins}
\usepackage{breqn}
\usepackage{slashed}
\usepackage{enumitem}
\usepackage[numbers,sort&compress]{natbib}
\usepackage{caption}
\usepackage{subcaption}
\usepackage[export]{adjustbox}

\graphicspath{ {figs/} }

\allowdisplaybreaks
\newcommand{\btoc}{b\to c\ell \bar{\nu}_\ell}
\newcommand{\btou}{b\to u\ell \bar{\nu}_\ell}

\begin{document}

\title{\vskip-3cm{\baselineskip14pt
    \begin{flushleft}
     \normalsize P3H-23-053, TTP23-030, ZU-TH 41/23, CERN-TH-2023-151
    \end{flushleft}} \vskip1.5cm
  Revisiting semileptonic $B$ meson decays at next-to-next-to-leading order
}
 
\author{
  Manuel Egner$^{a}$,
  Matteo Fael$^{b}$,
  Kay Sch\"onwald$^{c}$,
  Matthias Steinhauser$^{a}$
  \\
  {\small\it (a) Institut f{\"u}r Theoretische Teilchenphysik,
    Karlsruhe Institute of Technology (KIT),}\\
  {\small\it Wolfgang-Gaede Stra\ss{}e 1, 76128 Karlsruhe, Germany}
  \\
  {\small\it (b) Theoretical Physics Department, CERN, 1211 Geneva, Switzerland}
  \\
  {\small\it (c) Physik-Institut, Universit\"at Z\"urich, Winterthurerstrasse 190,}\\
  {\small\it 8057 Z\"urich, Switzerland}
}

\date{}

\maketitle

\thispagestyle{empty}

\begin{abstract}

  We compute next-to-next-to-leading order corrections to the
  semileptonic decay rate of $B$ mesons for arbitrary values of the
  final-state quark mass. For the contribution with one massive quark
  in the final state, we extend the literature result and obtain
  analytic expressions in terms of iterated integrals.  For the
  complete contribution, which also includes contributions with three
  massive quarks in the final state, we present a semi-analytic
  method, which leads to a precise approximation formula for the decay
  rate.  Our results agree with the expansions available for $\btoc$
  and $\btou$ in the literature. The main emphasize of this paper is
  on the technical aspects of the calculation which are useful for a
  wider range of applications.

\end{abstract}


\newpage


\section{Introduction}

The inclusive semileptonic $ B \to X \ell \bar \nu_\ell$ decays, mediated by
the charged-current transition $b \to q \ell \bar \nu_\ell$ with $q=u,c$, are
standard probes of the CKM matrix elements $V_{cb}$ and $V_{ub}$.  The
comparison between the experimental values of the branching ratios and their
theoretical predictions obtained within the framework of the Heavy Quark
Expansion has allowed the extraction of $|V_{cb}|$ with a 1.2\% accuracy
\cite{Bordone:2021oof,Bernlochner:2022ucr} while for $|V_{ub}|$ it has reached
about $5\%$~\cite{HeavyFlavorAveragingGroup:2022wzx}.

The calculation of the decay rate in the free quark approximation, 
i.e.\ $\Gamma(b \to q \ell \bar \nu_\ell)$, constitutes one of the main 
theoretical ingredients for the extraction of the CKM elements. 
The decay rate is known in an exact form for arbitrary
mass of the final state quark $q$ only up to order $\alpha_s$.
At higher orders, results have been obtained only as asymptotic expansions 
either in the massless limit, with the expansion parameter $\rho = m_c/m_b \ll 1$, or
the equal mass limit, with the expansion parameter $\delta = 1- \rho = 1 -m_c/m_b \ll 1$.

At next-to-next-to-leading order (NNLO), expansions around $\rho \to 0$,
which covers both $\btou$ and $\btoc$, have been computed in Refs.~\cite{Pak:2008qt,Pak:2008cp}. 
The asymptotic expansion in this limit is quite involved and it has not yet been extended to
next-to-next-to-next-to-leading order (N$^3$LO).
At NNLO also the other limit $\delta \to 0$ was studied in Ref.~\cite{Dowling:2008mc}
showing a much simpler asymptotic expansion (compared to the limit $\rho \to 0$)
and a fast convergence of the series in $\delta$ even at the physical value of $m_c$.  
At N$^3$LO an expansion around $m_c\simeq m_b$ has been performed in
Ref.~\cite{Fael:2020tow,Fael:2022frj} (see also Ref.~\cite{Czakon:2021ybq} where a subset
of diagrams have been cross-checked). 
Currently, the predictions for $b \to u$ at N$^3$LO are based on expansions for $m_c\to m_b$
and a subsequent extrapolation to $m_c=0$. This yields a sizable
uncertainty of about 10\%.
  
Recently, a semi-analytic method for the calculation of multi-loop 
Feynman integrals depending on one dimensionless parameter (and the dimension $d$) 
has been developed in Ref.~\cite{Fael:2021kyg}.
It was applied successfully in QCD to the calculation of the fermionic part
of the $\overline{\mathrm{MS}}$-pole mass relation at four loops and
the massive form factors at three loops~\cite{Fael:2022rgm,Fael:2022miw,Fael:2023zqr}.
The ``expand and match'' method provides results well suited for fast numerical evaluation 
and sufficiently precise for phenomenological applications.

In this paper we reconsider the corrections of $O(\alpha_s^2)$ to the
$b \to q \ell \bar \nu_\ell$ total rate, and revisit its calculation utilizing
the method developed in~\cite{Fael:2021kyg}.  The purpose is twofold. On the
one hand, the application of the ``expand and match'' method to semileptonic
decays serves as preparation for similar other calculations in $B$ physics,
such as the computation of non-leptonic decay rates at NNLO. It also provides
cross checks of the known expansions around $\rho = 0$ and $\rho = 1$.  On the
other hand, we discuss for the first time the role of the rare decay
$b \to c\bar{c}c \ell \bar{\nu}_\ell$, which contains an additional $\bar c c$ pair
in the final state. This decay channel was so far neglected in the expansions
around the limit $m_c \simeq m_b$, both at NNLO and N$^3$LO.

Let us further elaborate on the last point. At leading order and NLO, the
possible real emission processes which contribute to the total semileptonic
rate ($b \to c \ell \bar\nu_\ell$ and $b \to c g \ell \bar\nu_\ell$) contain
only one charm quark in the final state.  Starting from NNLO, there is also a
decay channel where three charm quarks appear in the final state (see, e.g.,
Figs.~\ref{fig::1c3c}(a) and (e)).

For technical reasons (see below) the calculation for $m_c \simeq m_b$ includes
only the contributions with one (massive) charm quark in the final state 
but neglects the contributions with three charm quarks. 
The latter is kinematically accessible only if $m_c<m_b/3$ 
and thus missing in the equal mass limit $m_c=m_b$. 
The contribution of the rare decay is very small,
of the order of $10^{-7}$, and so negligible for the physical value of the charm mass
($\rho \simeq 0.25$), without impact on the extraction of $V_{cb}$.
However, in order to properly match the expansion around $\rho =0$ and $\rho =1$
it is crucial to correctly take into account the contribution of $b \to c\bar cc \ell \bar \nu_\ell$.

The main results presented in this paper are the following:
\begin{itemize}
\item We compute analytic results for the NNLO correction in terms of iterated
  integrals for all contributions with one charm quark in the final state.  We
  do not include contributions with three charm quarks since they would
  involve elliptic integrals originating, e.g., from four-loop sunrise
  diagrams with unequal masses.
\item We construct a piece-wise defined function where the individual pieces
  are either Taylor or power-log expansions with (precise) numerical
  coefficients. This approximation contains all contributions, also those with
  three charm quarks in the final state.  It reproduces the correct functional
  behaviour in the various kinematic limits at $\rho \to0$ and $\rho \to1$ but
  also at $\rho \to 1/3$.

\end{itemize}

The outline of this paper is as follows: In the next Section we introduce the
notation and describe our analytic and numeric methods. Afterwards we discuss
in Section~\ref{sec::decay} our results for the total decay rate of $b \to c \ell \bar \nu_\ell$
at NNLO. In particular, we present analytic results for the contributions with one and
three charm quarks in the final state. We also discuss our approximation
formulas and in particular their numerical accuracy. 
In Section~\ref{sec:Uc} we present the calculation of the charm dependent contribution to the
decay $b \to u \ell \bar \nu_\ell$, i.e.\ the two-loop diagrams 
with a closed charm loop insertion into the gluon propagator.
Finally we summarize our findings in Section~\ref{sec::concl}. 


\section{\label{sec::method}Methods}


\subsection{\label{sub::notation}Notation}
Let us now discuss the details of the calculation.
We consider the decay of an on-shell bottom quark into
a charged lepton ($e, \mu$) assumed to be massless, a neutrino
and any hadronic state $X_c$ containing a charm quark:
\begin{equation}
  b \to X_c \ell \bar \nu_\ell.
\end{equation}

To compute the decay rate we apply the optical theorem
and compute the imaginary part of the bottom quark two-point function.
In the following we do not restrict ourselves to physical masses of $m_c \approx
m_b/3$, rather we keep $m_c$ generic.  As we will see, for technical
reasons the limit $m_c\to m_b$ is of high relevance. 
With $m_c\to 0$ we cover the case $\btou$.  
After specifying the QCD colour factors to
QED, our results can also be applied to the muon and tau decays.

It is convenient to define the following variables:
\begin{align*}
  \rho &= \frac{m_c}{m_b}\,, &
  \delta &= 1- \frac{m_c}{m_b}\,.
\end{align*}
We write the decay rate in the form
\begin{eqnarray}
  \Gamma(B \to X_c \ell \bar\nu) 
  &=& \Gamma_0 \left[X_0 + C_F\sum_{n\ge 1}
      \left(\frac{\alpha_s}{\pi}\right)^n X_n
      \right] 
      + {\cal O}\left(\frac{\Lambda_{\rm QCD}^2}{m_b^2}\right)\,,
      \label{eq::gamb2c}
\end{eqnarray}
where $m_b$ and $m_c$ are renormalized in the on-shell scheme,
\begin{eqnarray}
  \Gamma_0 &=& \frac{ A_{\rm ew}G_F^2|V_{cb}|^2 m_b^5 }{ 192\pi^3 }
               \,,
\end{eqnarray}
$C_F=4/3$, $A_{\rm ew}=1.014$ is the leading electroweak
correction~\cite{Sirlin:1981ie}, and $\alpha_s\equiv\alpha_s^{(5)}(\mu_s)$
with $\mu_s$ being the renormalization scale.
The tree-level contribution is given by
\begin{eqnarray}
  X_0 &=& 1 - 8\rho^2 - 12\rho^4 \log(\rho^2) + 8\rho^6 - \rho^8
          \,.
\end{eqnarray}
The one- and two-loop results are available from
Refs.~\cite{Trott:2004xc,Aquila:2005hq,Pak:2008qt,Pak:2008cp,Melnikov:2008qs,Biswas:2009rb,Gambino:2011cq,Dowling:2008mc}
and the three-loop correction $X_3$ has been computed in
Ref.~\cite{Fael:2020tow} (see also~\cite{Czakon:2021ybq} for partial results).
In the following we reconsider $X_2$ which is a function of the mass ratio $m_c/m_b$. 

\begin{figure}[t]
  \centering
  \begin{tabular}{ccc}
    \includegraphics[width=0.3\textwidth]{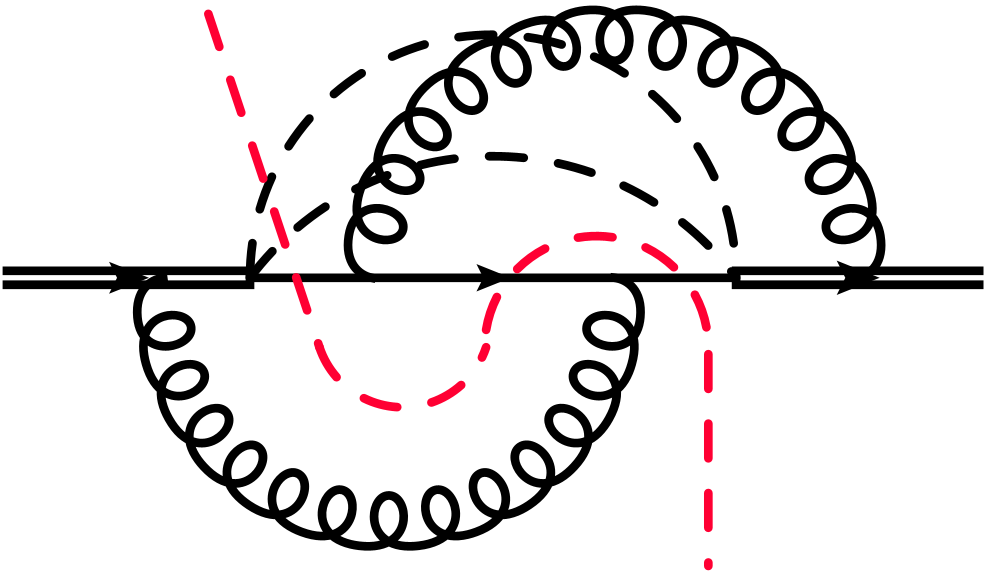} &
    \includegraphics[width=0.3\textwidth]{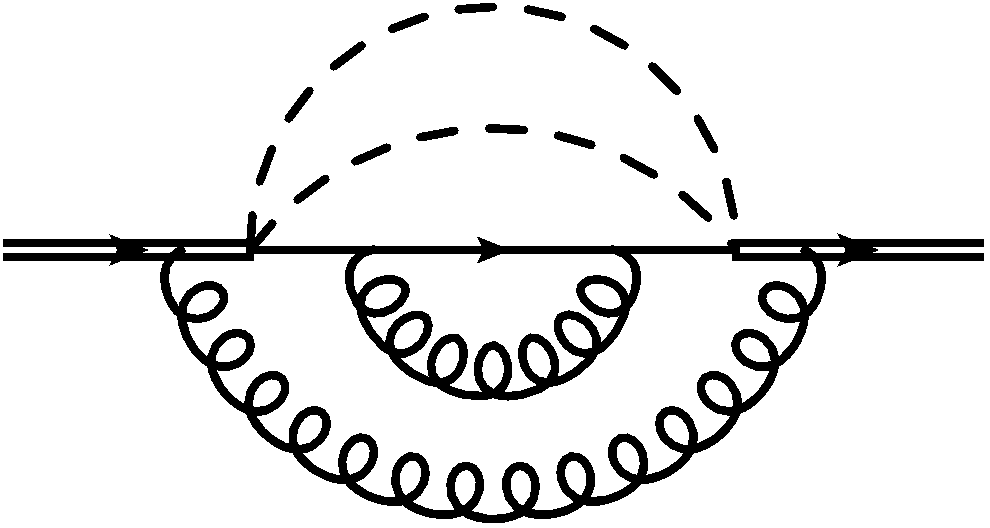} &
    \includegraphics[width=0.3\textwidth]{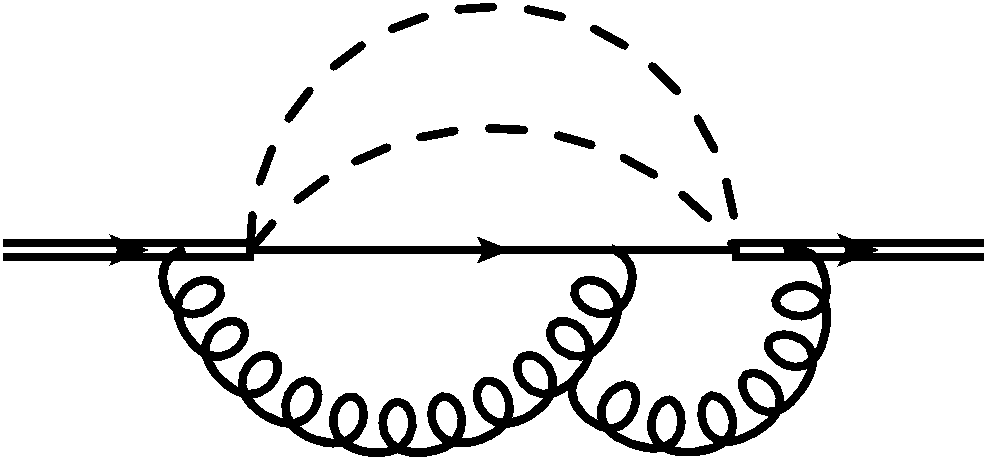}
    \\ (a) & (b) & (c) \\[1em]
    \includegraphics[width=0.3\textwidth]{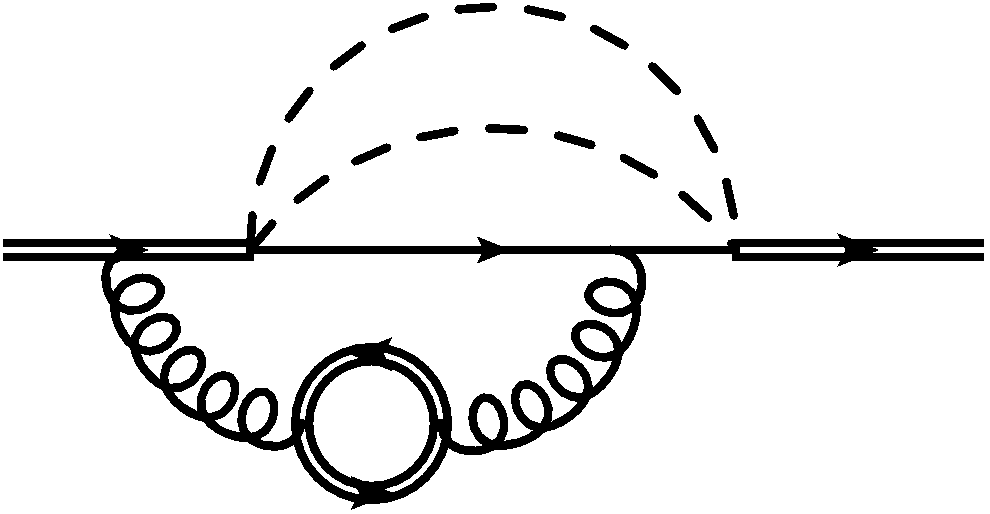} &
    \includegraphics[width=0.3\textwidth]{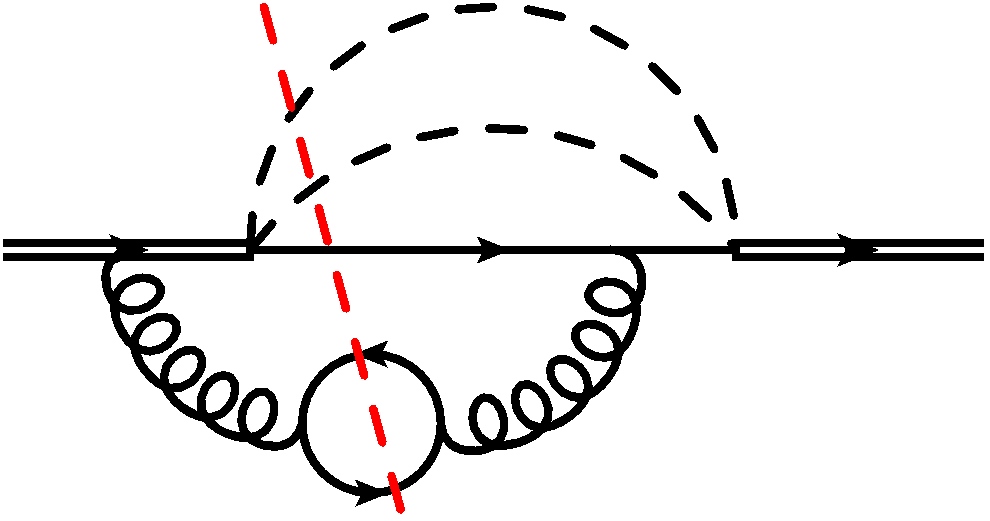} &
    \includegraphics[width=0.3\textwidth]{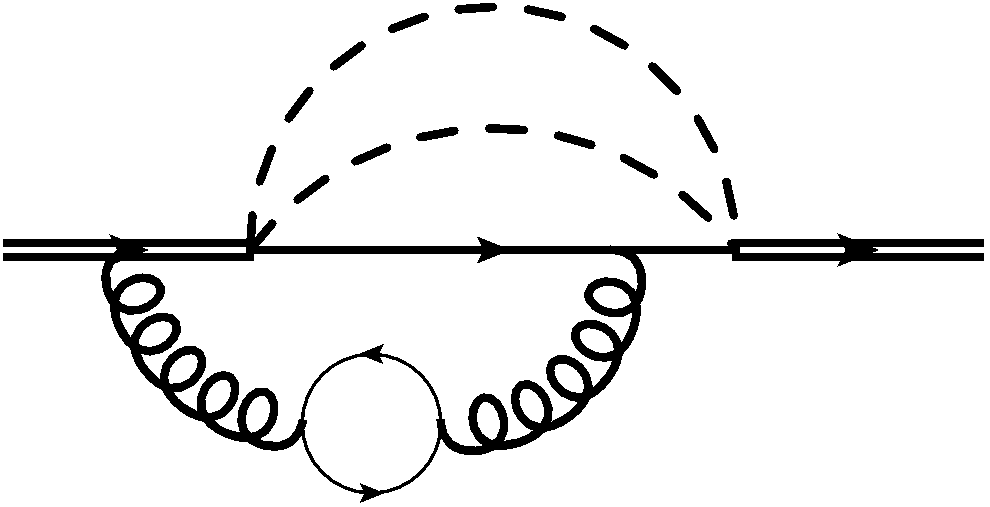}
    \\ (d) & (e) & (f) \\[1em]
    \includegraphics[width=0.3\textwidth]{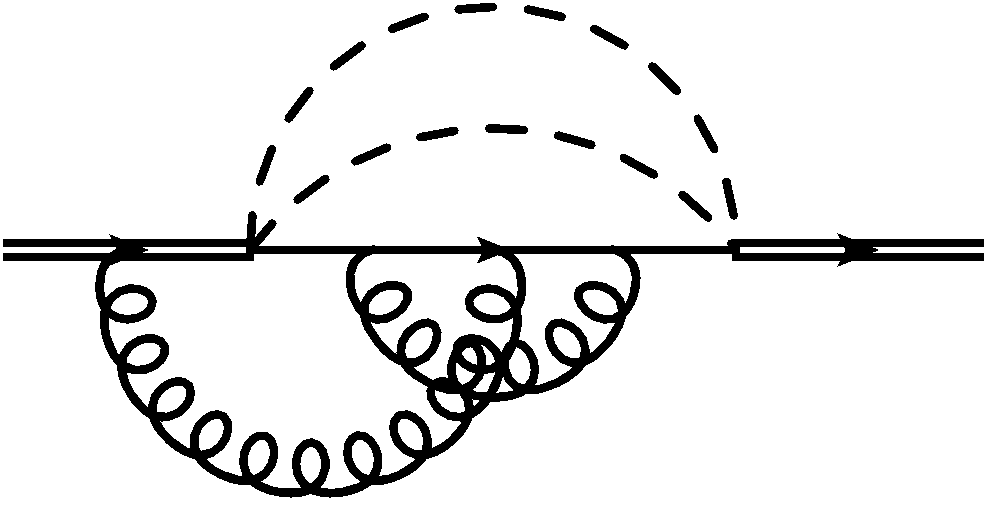} &
    \includegraphics[width=0.3\textwidth]{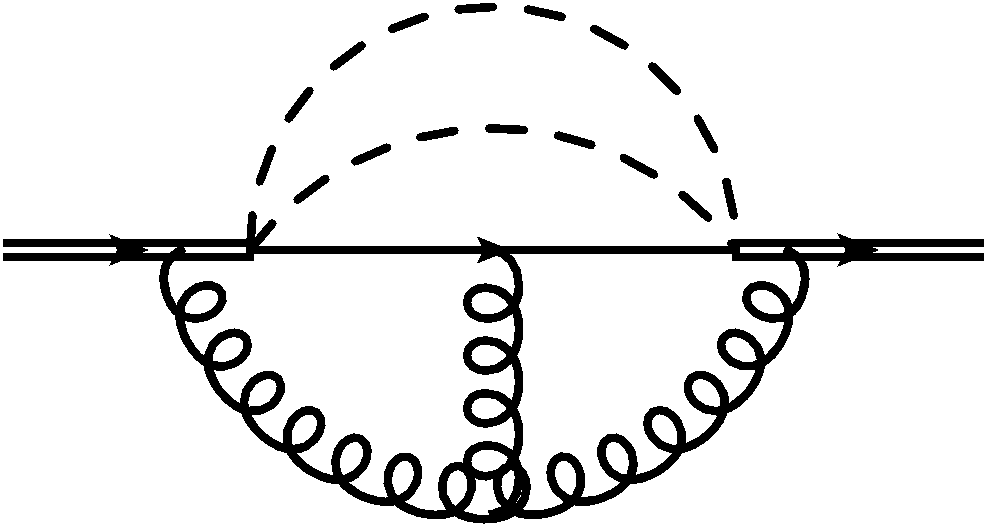} &
    \includegraphics[width=0.3\textwidth]{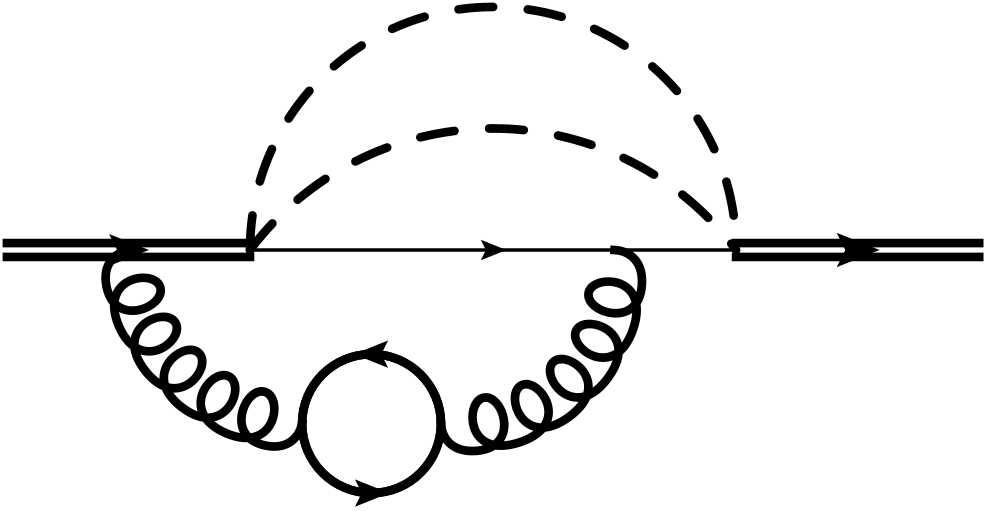}
    \\ (g) & (h) & (i)
  \end{tabular}
  \caption{\label{fig::1c3c}Sample diagrams contributing to the semileptonic
    $B$ meson decays at NNLO. The dashed lines represent the charged lepton
    neutrino pair. Double lines denote bottom quarks, thick and thin lines
    denote charm and massless quarks, respectively. The red dashed lines
    shown for diagrams (a) and (e) indicate possible three-charm cuts.}
\end{figure}
At leading order and NLO, the functions $X_0$ and $X_1$ come
from Feynman diagrams with cuts only through one charm quark line.
These diagrams develop an imaginary part for $0 \le \rho < 1$.
Starting at NNLO, there are also diagrams where three charm lines
can be cut (see, e.g., Fig.~\ref{fig::1c3c}(a) and (e)). 
They develop an additional discontinuity for $0 \le \rho < 1/3$ which corresponds
to the contribution of the rare decay $b \to c\bar{c}c \ell \nu$.

In the following we present our calculation of $X_2$ which is divided in two steps.
In a first step, we compute analytic results valid for arbitrary charm quark masses 
for the contribution to $X_2$ which does not contain three charm quarks in the final state.  
Later on, we present an approximation function for the complete
contribution to $X_2$, which consists of power-log expansions valid in certain
$m_c$ ranges.


\subsection{\label{sub::setup}General setup}

We generate the amplitude with {\tt qgraf}~\cite{Nogueira:1991ex} and process
the output with {\tt tapir}~\cite{Gerlach:2022qnc}. Next we apply {\tt
  exp}~\cite{Harlander:1997zb,Seidensticker:1999bb} to map the individual
diagrams to integral families.  The actual computation is done with {\tt
  FORM}~\cite{Ruijl:2017dtg}.  At this step the auxiliary files generated by
{\tt tapir} are quite useful to express the amplitude for each diagram as a
linear combination of scalar integrals.

They are reduced to master integrals with the use of {\tt  Kira}~\cite{Klappert:2020nbg}
in combination with {\tt  FireFly}~\cite{Klappert:2020aqs} within each integral family. 
Furthermore, we apply {\tt ImproveMasters}~\cite{Smirnov:2020quc} in order to obtain for
each integral family a basis such that the $\epsilon$ and $\rho$ dependence in the
denominators of each reduction table entry factorizes. Once all master
integrals are identified we use again {\tt Kira} to identify the symmetries across
families and reduce the master integrals to a minimal set. 
In total we arrive at 129 four-loop master integrals. 
Samples of master integrals are shown in Fig.~\ref{fig:MI}

Note that we do not exploit that it is in principle possible to integrate over
the charged-lepton-neutrino loop in a first step since this cannot be done for
the calculation of the non-leptonic $B$ meson decays.

We perform the calculation for general QCD gauge parameter $\xi$. For most of
the colour structures $\xi$ drops out at the level of the bare amplitude. The
remaining $\xi$ dependence cancels after the inclusion of the bottom mass
counterterms.

A crucial input for the methods discussed in the next two subsections
are the differential equations for the master integrals. They can easily
be established using the reduction tables.


\subsection{\label{sub::ana_calc}Analytic calculation}
\begin{figure}[t]
  \centering
  \centering
  \begin{tabular}{ccc}
    \includegraphics[valign=c,width=0.2\textwidth,clip,trim={4.5cm 0 4.5cm 0}]{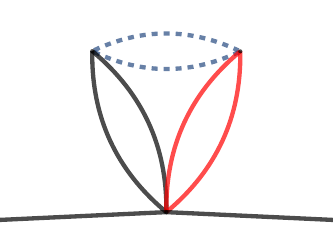}&
    \includegraphics[valign=c,width=0.25\textwidth,clip,trim={3.5cm 0 3.5cm 0}]{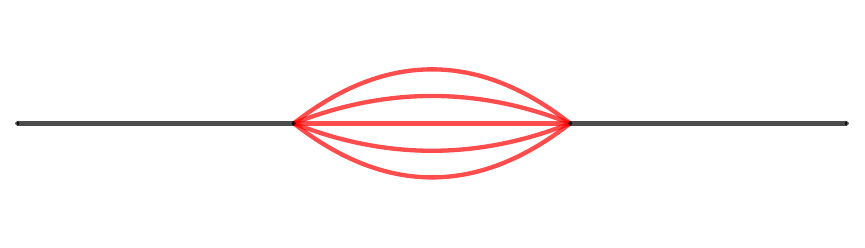}&
    \includegraphics[valign=c,width=0.25\textwidth,clip,trim={3.5cm 0 3.5cm 0}]{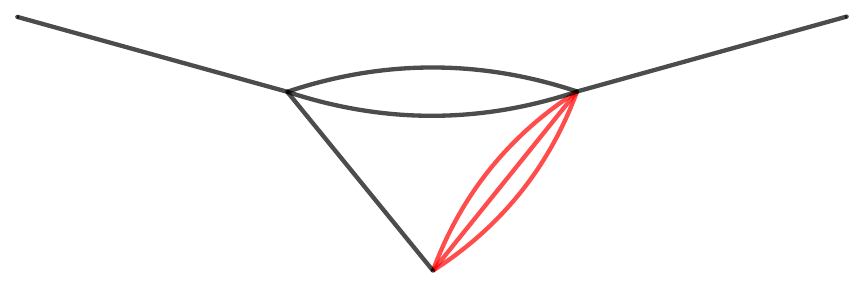}
    \\
    \includegraphics[valign=c,width=0.25\textwidth,clip,trim={3.5cm 0 3.5cm 0}]{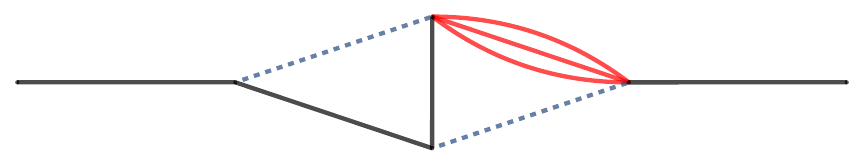}&
    \includegraphics[valign=c,width=0.25\textwidth,clip,trim={3.5cm 0 3.5cm 0}]{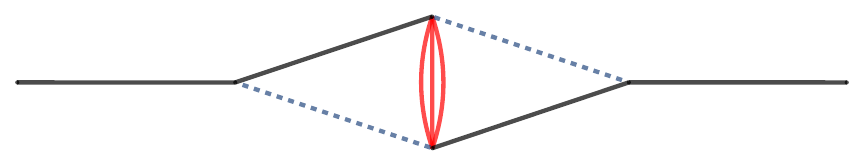}&
    \includegraphics[valign=c,width=0.25\textwidth,clip,trim={3.5cm 0 3.5cm 0}]{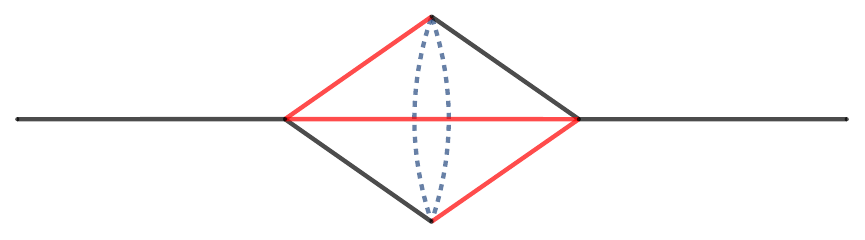}
    \\
    \includegraphics[valign=c,width=0.25\textwidth,clip,trim={3.cm 0 3.cm 0}]{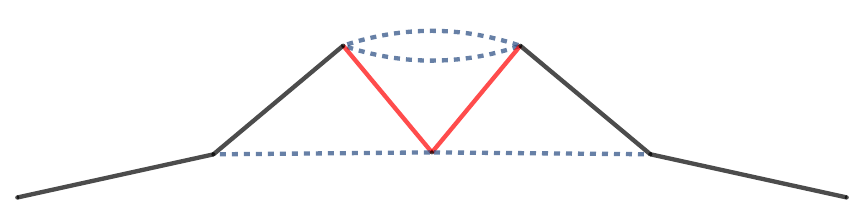}&
    \includegraphics[valign=c,width=0.25\textwidth,clip,trim={3.5cm 0 3.5cm 0}]{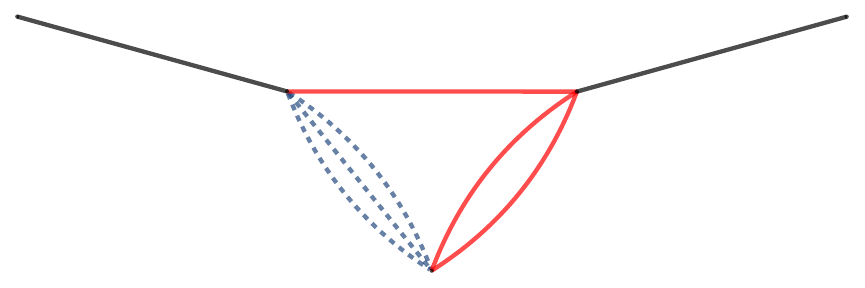}&
    \includegraphics[valign=c,width=0.25\textwidth,clip,trim={3.cm 0 3.cm 0}]{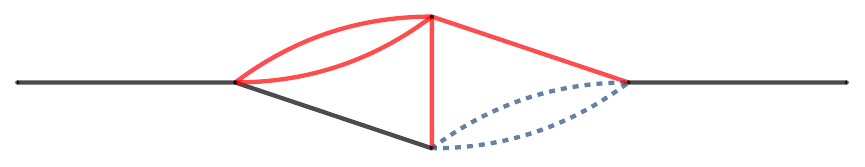}
  \end{tabular}
  \caption{Samples of four-loop master integrals. Black and red solid lines represents
  massive propagator with mass $m_b$ and $m_c$, respectively, while dashed lines are
  massless propagators.}
  \label{fig:MI}
\end{figure}
In this subsection we present our analytic calculation of the master
integrals, in case we consider the contribution of cuts only through one charm
quark.  For the decay rate, we need in principle only the imaginary parts of
the master integrals.  As a consequence we can restrict ourself only to the
masters which have a physical cut, i.e.\ where the propagator of neutrino,
charged lepton and at least one charm propagator can be cut simultaneously.
In Fig.~\ref{fig:MI}, the master integrals in the first row do not have a
physical cut, while those in the second and third row have an imaginary part.

The requirement of physical cuts reduces the number of master integrals from
  129 to 108.
We construct analytic solutions for the imaginary part of the master integrals 
with the help of the differential equations. 
We use boundary conditions from the limit $m_c\to m_b$ ($\rho\to1$) where it
is quite straightforward to compute the imaginary parts of the
master integrals via an asymptotic expansion.

The crucial observation is that by fixing the boundary conditions at
$m_c = m_b$ one calculates the contribution where only one charm line
is cut.  In fact, in this limit there is no possible discontinuity
corresponding to a cut through three charm quarks, which appears in
Feynman integrals only for $0 \le \rho<1/3$ (i.e.\ $\delta>2/3$).  For
example, the master integrals shown in the second row of
Fig.~\ref{fig:MI} have a cut only through three charm lines
(additionaly to the lepton and neutrino), while those in the third row
have a cut through one charm line.  Note that for some of the masters,
like the one on the right in the third row, both one-charm and
three-charm cuts appear.

In order to incorporate the three-charm cut contribution, it would be
necessary to compute a complicated asymptotic expansion around $\rho =
1/3$.  Alternatively one can compute boundary conditions for both real
and imaginary parts for $\rho>1/3$ and solve the differential
equations for the complete set of 129 master integrals.  This would
require the computation of four-loop on-shell integrals, which are to
date not available in analytic form.  The semi-analytic approach
discussed in Section~\ref{sub::numeric} covers this case.  In the
limit $m_c\to m_b$ only 95 out of the 129 master integrals have a cut
through one charm line.  Note that 13 master integrals have cuts only
through 3 charm lines while the remaining 21 do not have an imaginary
part.

Let us now discuss our analytic solution for the 95 masters with one
charm cut.  In a first step we transform the differential equation matrix in
$\epsilon$ form~\cite{Henn:2013pwa,Lee:2014ioa}.  This is done using both {\tt
  Libra}~\cite{Lee:2020zfb} and {\tt Canonica}~\cite{Meyer:2017joq}.

{\tt Libra} is used to bring the matrix in block-diagonal form and afterwards
we use {\tt Canonica} to transform the whole system to $\epsilon$ form.  This
is done by first transforming the diagonal blocks and then the corresponding
off-diagonal elements block by block, using the build-in functions of {\tt
  Canonica}.  This approach is successful for a subset of 91 master integrals.
The remaining four integrals are at the top-level and can be decomposed into a
$3 \times 3$ system and an uncoupled integral.  We solve these integrals
following the algorithm outlined in Ref.~\cite{Ablinger:2018zwz}.  In practice
this means that we decouple the coupled system with the package {\tt
  OreSys}~\cite{ORESYS}, which internally depends on {\tt
  Sigma}~\cite{Schneider:2007}, and solve the resulting higher order
differential equation via factorization of the differential operator with the
help of {\tt HarmonicSums}~\cite{HarmonicSums}.  In an independent calculation
we have used this approach of solving the differential equation on the whole
system and did not make use of the $\epsilon$ form of the first 91 master
integrals.  The results of both approaches are in complete agreement.

The $\epsilon$ form is conveniently obtained in the variables $t$
which is defined via
\begin{eqnarray}
  \rho &=& \frac{1-t^2}{1+t^2}\,,
  \nonumber\\
  t &=& \frac{\sqrt{1-\rho}} {\sqrt{1+\rho}}\,.
\end{eqnarray}
Then the solution can be expressed in terms of iterated integrals
with the alphabet
\begin{equation}
  \left\{ \frac{1}{1+t},\frac{1}{t},\frac{1}{1-t},
          \frac{t}{1+t^2},\frac{t^3}{1+t^4} \right\}.
	\label{eq::alphabet}
\end{equation}

For the computation of the boundary conditions we follow
Refs.~\cite{Fael:2020njb,Fael:2020tow,Fael:2022frj}, where similar integrals have been
considered.  In~\cite{Fael:2020iea,Fael:2020njb} the three-loop relation between the heavy
quark masses defined in the pole and kinetic scheme has been
computed. The starting point in Refs.~\cite{Fael:2020tow,Fael:2022frj} were
five-loop integrals. A convenient choice of momentum routing and  integration order
led to a factorization, where at most three-loop integrals have to be
solved. In the present calculation we deal with four-loop integrals
and observe a similar factorization.

We apply the so-called ``method of regions''~\cite{Beneke:1997zp} which in our
case leads to a scaling of the loop momenta
as either hard ($h$) or ultra-soft ($us$).\footnote{Following Ref.~\cite{Beneke:1997zp} we use the term
ultra-soft instead of soft.}  This leads to at most $2^4=16$
different regions for a given integral, however, not all of them
contribute. For example, the region where all loop momenta are hard
does not develop an imaginary part and can be discarded.

For the identification of the regions we use the
package {\tt asy.m}~\cite{Pak:2010pt}, which we apply to each master integral
separately. As input it requires the list of propagators and the
scalings of the masses and the external momenta. {\tt asy.m} provides
as output the scalings of the propagators for all non-vanishing
regions. The comparison to the scalings which we obtain after
identifying a unique routing of the loop momenta allows us to discard
regions, which give no contribution. For the non-vanishing regions we
can perform the expansions in the small parameter $\delta$.

Note that in Refs.~\cite{Fael:2020njb,Fael:2020tow}, the program \texttt{asy.m}
was only used as a cross check to make sure that all contributing
regions were found. It was furthermore applied at the level of the
amplitudes and not to master integrals. In the present calculation
\texttt{asy.m} is used to discard regions which are zero before the
expansion is done. Since we use it at the level of the master
integrals, it is suitable to apply this procedure to each of the 95
master integrals separately.

Two of the four loop momenta have to be ultra-soft, namely the loop
momentum of the charged-lepton-neutrino loop and the one which flows
into the lepton loop. The remaining two loop momenta are either ultra-soft
or hard. 

The calculation of the Feynman integrals appearing from the asymptotic
expansion closely follows Ref.~\cite{Fael:2022frj} and can be summarized by
the following steps.  We consider separately the regions corresponding to the
momentum scaling $(h,h,us,us), (h,us,us,us)$ and $(us, us,us,us)$.  For each
region we define new integral families that contain all propagators and
numerators which arise after the expansion.  We find symmetries across the new
integral families with the program {\tt LIMIT}~\cite{Herren:2020ccq}, which is
based on {\tt LiteRed}~\cite{Lee:2013mka}.  In case there are linearly
dependent propagators, we perform a partial fraction decomposition with {\tt
  LIMIT} and minimize again the number of families.  For each family, we
perform a reduction to master integrals with {\tt
  Kira}~\cite{Klappert:2020nbg}.  The master integral in the asymptotic
expansion can be written in terms of $\Gamma$ functions and, in one case, as a
one-fold Mellin-Barnes integral.  In particular, the master integrals with
three or four ultra-soft loop momenta can be solved via recursive one-loop
integration.  In case we have two hard loop momenta, we observe a
factorization into one- and two-loop integrals.

For convenience we provide the formulas for one-loop integrals used in the 
recursive integration (with $q^2=1$):
\begin{align}
  I_1(p; n_1,n_2) &= 
  \int  \frac{\mathrm{d}^dk_1}{\left(-k_1^2\right)^{n_1} 
  \left(-\left(k_1+p\right)^2\right)^{n_2} } \nonumber \\ 
  & = 
  i \pi^{d/2}
  \frac{\Gamma\left(n_1+n_2-2+\epsilon\right) 
  \Gamma\left(2-n_2-\epsilon\right) 
  \Gamma\left(2-n_1-\epsilon \right)}
  {\Gamma\left(n_1\right) 
  \Gamma\left(n_2\right)
  \Gamma\left(4-n_1-n_2-2\epsilon\right)}
  \left( -{p^2} \right)^{2-\epsilon+n_1+n_2}
  , \nonumber \\
  I_2 (n_1,n_2)&= 
  \int \frac{\mathrm{d}^dk_1}{ \left( -k_1^2 \right)^{n_1}
  \left(-k_1^2+2 k_1 \cdot q \right)^{n_2} } \nonumber \\ 
  & = i\pi^{d/2}
  \frac{\Gamma\left(n_1+n_2-2+\epsilon \right)
  \Gamma\left(4-n_2-2n_1-2\epsilon \right)}
  {\Gamma\left( n_2\right)
  \Gamma\left( 4-n_1-n_2-2\epsilon\right) }, \nonumber \\
  I_3(\delta; n_1,n_2,n_3) &= 
  \int \frac{\mathrm{d}^dk_1}{\left(-k_1^2\right)^{n_1}
                     \left(-2q \cdot k_1\right)^{n_2}
                     \left(-\delta-2q \cdot k_1\right)^{n_3}} \nonumber\\ 
  & =
  i \pi^{d/2}
  \left(-\delta\right)^{4-2\epsilon-2n_1-n_2-n_3} \nonumber\\&
  \qquad \frac{\Gamma\left(4-2\epsilon-2n_1-n_2\right)\Gamma\left(2-\epsilon-n_1\right)
  \Gamma\left(2n_1+n_2+n_3-4+2\epsilon\right)}
  {\Gamma\left(n_1\right)\Gamma\left(n_3\right)\Gamma\left(4-2\epsilon-2n_1\right)}.
  \nonumber
\end{align}
In the hard regions, we used also the expression for the two-loop
sunrise diagram with three equal masses:
\begin{align}
  I_4 \left(1,1,1\right) &= 
  \int \frac{\mathrm{d}^dk_1\mathrm{d}^dk_2}{
  \left(1-\left(k_1+k_2\right)^2\right) 
  \left(1-\left(k_1\right)^2\right) 
  \left(1-\left(k_2+q\right)^2\right)} \nonumber\\
   &=
   \left( i \pi^{d/2} \right)^2
   \frac{1}{2 \pi i} \int_{-i \infty}^{i \infty} \mathrm{d}z  \nonumber\\&
    \qquad\frac{\Gamma\left(-z\right)
    \Gamma^2\left(1-\epsilon-z\right) 
    \Gamma\left(-1+2\epsilon+z\right) 
    \Gamma\left(1-\epsilon\right)
    \Gamma\left(2-2\epsilon-z\right)}
    {\Gamma\left(2-2\epsilon-2z \right) 
    \Gamma\left(3-3\epsilon-z\right)\left(-4\right)^{-1+2\epsilon+z}}.
     \label{eq::MIs}
\end{align}

We compute for each master integral the first two expansion
coefficients for $\delta\to 0$.  Note that not all $\delta$
coefficients are needed to fix the integration constants.  There are
diagonal blocks in the differential equation where several integrals
are coupled. It is usually sufficient to choose one of the integrals
and match it to its boundary condition including subleading $\delta$
terms.  This allows us to fix all boundary conditions also for the remaining
integrals in the block.  Once all integration constants are fixed, we
calculate from the analytic solution the $\delta$ expansion of all
integrals. We use the coefficients which were not used as cross
check. Altogether we have computed 192 coefficients in the $\delta$
expansion (two for each of the 95 master integrals) but we have used
only 72 of them to fix the boundary constants.

Once all boundary conditions are fixed it is straightforward to integrate the
differential equations up to the required order in $\epsilon$.  After
inserting the master integrals into the amplitude, we expand in $\epsilon$ up
to the constant term. The bare NNLO amplitude develops $1/\epsilon^2$ and
$1/\epsilon$ poles which cancel against the counterterm contribution. Their
transcendental weight is one and two, respectively.  In the $\epsilon^0$ term
we observe iterated integrals up to weight five even after choosing a minimal
set of functions.
The occurrence of weight-two expressions in the pole part is expected since
this weight also appears in the analytic NLO result~\cite{Nir:1989rm}.
At first sight the weight-five functions in the finite three-loop term
might be surprising. Note, however, that we start with
four-loop integrals. Since the charged-lepton-neutrino loop is finite
one can expect weight-six expressions from the remaining three-loop
calculation. Since we compute the imaginary parts we finally end up with
weight-five iterated integrals.

We refrain from presenting exact analytic results in the paper but refer to
the supplementary material~\cite{progdata} where also the expressions for the
imaginary part of the master integrals (one charm cut) can be found.  In
Section~\ref{sub::ana_exp} we will present the analytic expansions of the
decay rate in the limits $\rho\to0$ and $\rho\to1$.


\subsection{\label{sub::numeric}Numeric calculation}

As an alternative to the analytic approach described in the previous
subsection, we discuss in the following the ``expand and match''
method introduced in Refs.~\cite{Fael:2021kyg,Fael:2022miw}.  The starting point is the
system of differential equations for the 129 master integrals obtained
in Section~\ref{sub::setup}.  The basic idea of this method is as
follows: We choose several points $\rho_0\in[0,1]$, make an ansatz
for the expansion of the master integrals around $\rho=\rho_0$,
insert the ansatz in the differential equations and solve the
resulting system of linear equations for the coefficients in the ansatz
in terms of a few initial values. The latter are determined by
matching to known results.

The choice of expansion points $\rho_0$ should include all
singular values of the differential equations that
in our case are $\rho =0,1/3$ and $1$, where $\rho=1/3$ 
corresponds to the three-charm threshold.  
Furthermore we add regular points to obtain higher
precision approximation formulas. 
As expansion points we choose
\begin{eqnarray}
  \rho_0\in\{0,1/12,1/6,1/4,1/3,1/2,1\}
  \,,
  \label{eq::delp_reg}
\end{eqnarray}
with $\rho_0=1/2$ as starting point.  Note that the radius of convergence in
general extends only up to the next singular point in the complex plane. 
As a consequence it is sufficient to choose only a few points above the threshold at $\rho=1/3$.
Below threshold more expansion points are needed
in order to reach a good convergence.
Furthermore we employ M{\"o}bius transformations (see
Ref.~\cite{Lee:2017qql}) to extend the radius of convergence
of the series expansions into the direction of the farther
singularity. Explicit ready-to-use formulas for our
application can be found in Ref.~\cite{Fael:2022miw}.

The expansion around regular points is a simple Taylor expansion
and thus we choose the following ansatz for the master integrals
\begin{eqnarray}
  I_i &=&
  \sum_{j=\epsilon_{\mathrm{min}}}^{\epsilon_{\mathrm{max}}}
  \sum_{n=0}^{n_{\mathrm{max}}}
  c_{i,j,n} \epsilon^j \left(\rho - \rho_0 \right)^n
  \,,
  \label{eq::TaylorAnsatz}
\end{eqnarray}
where $\epsilon_{\mathrm{min}}$ is determined by the highest pole and
$\epsilon_{\mathrm{max}}$ depends on the spurious poles in the
amplitude in front of the respective master integral. For the expansion
depth we typically choose $n_{\rm max}=50$ which provides
about ten or more significant digits in our final result.

For the singular points the ansatz has to be extended to allow for logarithmic terms.
At $\rho_0=0$ we have
\begin{eqnarray}
  I_i &=&
  \sum_{j=\epsilon_{\mathrm{min}}}^{\epsilon_{\mathrm{max}}}
  \sum_{m=0}^{j+4}  \sum_{n=0}^{n_{\mathrm{max}}}
  c_{i,j,m,n} \epsilon^j \, \rho^n \, \log^m\left(\rho\right)\,.
  \label{eqLLAnsatzdel0}
\end{eqnarray}
At $\rho_0=1/3$ and $\rho_0=1$ our ansatz is
\begin{eqnarray}
  I_i &=&
  \sum_{j=\epsilon_{\mathrm{min}}}^{\epsilon_{\mathrm{max}}}\sum_{m=0}^{j+4}
  \sum_{n=n_{\mathrm{min}}}^{n_{\mathrm{max}}}
  c_{i,j,m,n} \epsilon^j \,  (\rho -\rho_0)^n \, 
   \log^m\left(\rho-\rho_0\right)\,.
\label{eq::powerlog}
\end{eqnarray}
We choose $n_{\rm min}<0$ to allow for potential negative powers. However, it
turns out that only the coefficients for $n\ge0$ are different from zero.
In case our ansatz is insufficient the system of linear equations
can either not be solved or only the trivial solution is possible. 

The expansion around the three-particle threshold at $\rho=1/3$ does
not require the introduction of half-integer powers of $\rho - 1/3$,
at variance with what is observed for two- and four-particle
thresholds (e.g., see \cite{Fael:2022rgm,Fael:2022miw}).  In fact the
production close to threshold of $n$ particles of mass $m$ behaves as
\cite{Davydychev:1999ic,kmelnikov}
\begin{equation}
  (s-(n\, m)^2)^{\frac{3n-5}{2}} \,,
  \label{eq::v_pow}
\end{equation}
where $s$ is the squared energy available to the system. 
One observes that half-integer powers are arising only when an even number
of particles are produced, while for an odd number the exponent is an integer.

As boundary condition for the ``expand and match'' procedure we choose
$\rho_0=1/2$. We use \texttt{AMFlow}~\cite{Liu:2022chg} to obtain
numerical results for all master integrals (both real and imaginary
part) with a precision of 80 digits and fix undetermined
constants for the Taylor expansion around $\rho = 1/2$ (see
Eq.~\eqref{eq::TaylorAnsatz}).  Next we evaluate the master integrals
at $\rho_0=0.4$ which serves as input for the expansion around
$\rho_0=1/3$. We could proceed in a similar way towards
$\rho=1$. However, here we can use the analytic boundary conditions to
produce a deep expansion in $1-\rho$.

At $\rho_0=1/3$ we perform the matching at a value $\rho>1/3$. Thus the
logarithms in Eq.~(\ref{eq::powerlog}) are real-valued whereas the
coefficients $c_{i,j,m,n}$ have both real and imaginary parts. After crossing
the threshold to values of $\rho<1/3$ the logarithms develop additional
imaginary parts according to
$\log(\rho-\rho_0) = \log(|\rho-\rho_0|) - i \pi$
which arise from the three-charm threshold.  For the expansion
around $\rho_0=1$ it is convenient to use the ansatz in
Eq.~(\ref{eq::powerlog}) with $(\rho -\rho_0)$ replaced by $(\rho_0 -\rho)$
since we always have $\rho<\rho_0$ and thus we do not have to deal with
spurious imaginary parts.  In this way, we determine the coefficients of all
expansions around the singular and regular points given in
Eq.~(\ref{eq::delp_reg}).

\begin{figure}[t]
  \centering
  \begin{tabular}{c}
    \includegraphics[width=.9\textwidth]{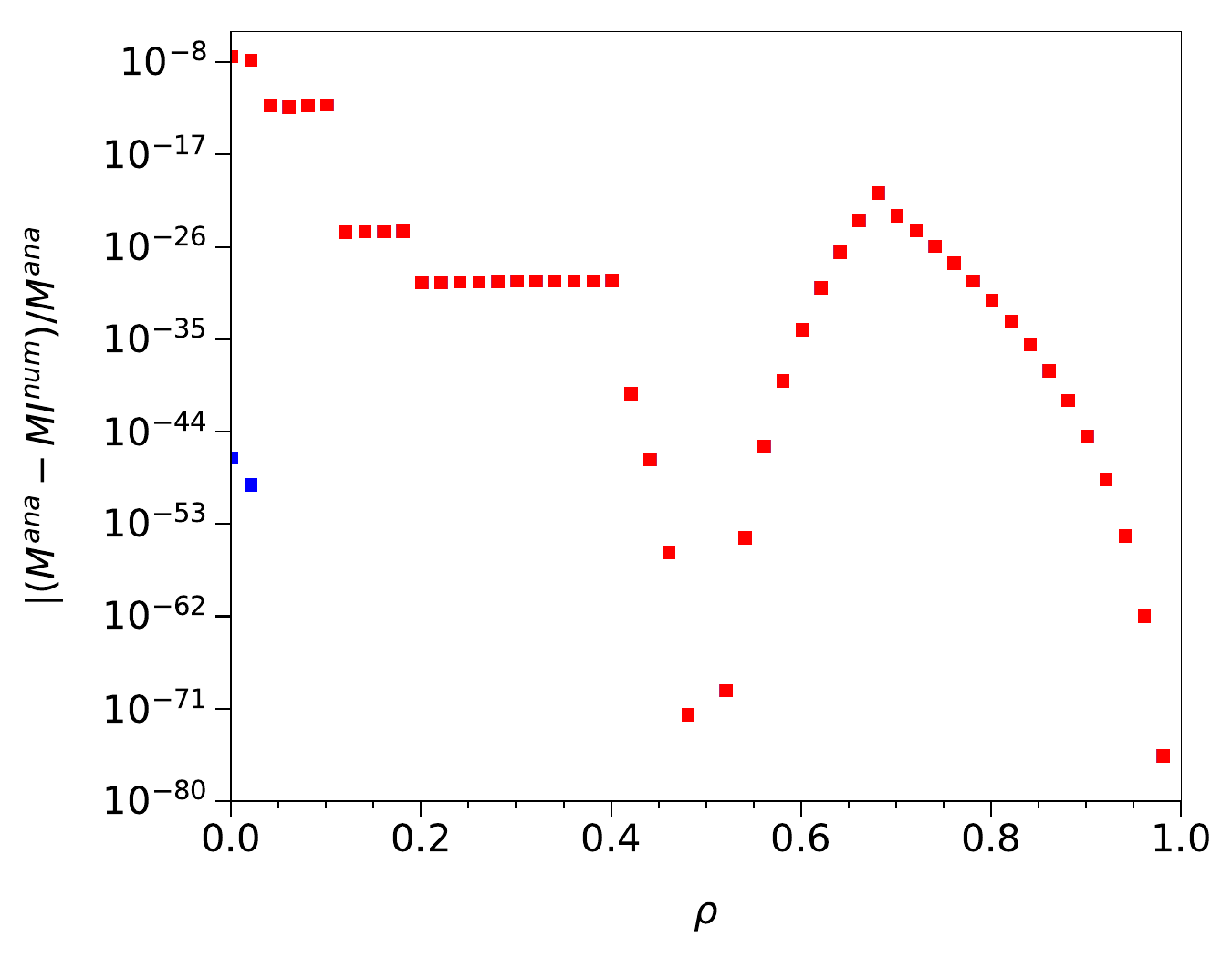}
  \end{tabular}
  \caption{\label{fig::compare_analytic}Relative difference between the
    semi-analytic approximation formula and the analytic result for the
    $\epsilon^0$ term of the (imaginary part) of the fourth non-planar master
    integral shown in Fig.~\ref{fig::compare_mi}.  The blue dots are
      obtained from the matching of the $\rho=0$ expansion to boundary
      conditions computed at $\rho=0.01$.}
\end{figure}

\begin{figure}[t]
  \centering
  \begin{tabular}{cccc}
    \includegraphics[valign=c,width=0.22\textwidth,clip,trim={3.cm 0 3.cm 0}]{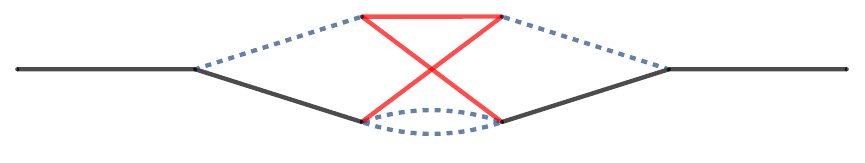}&
    \includegraphics[valign=c,width=0.22\textwidth,clip,trim={3.cm 0 3.cm 0}]{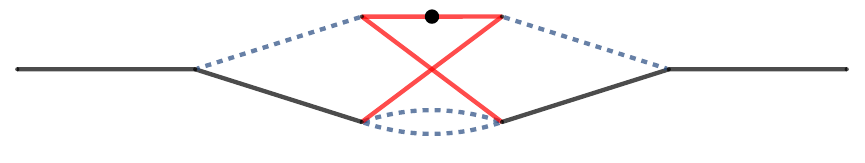}&
    \includegraphics[valign=c,width=0.22\textwidth,clip,trim={3.cm 0 3.cm 0}]{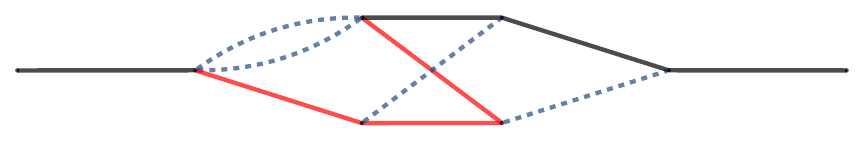}&
    \includegraphics[valign=c,width=0.22\textwidth,clip,trim={3.cm 0 3.cm 0}]{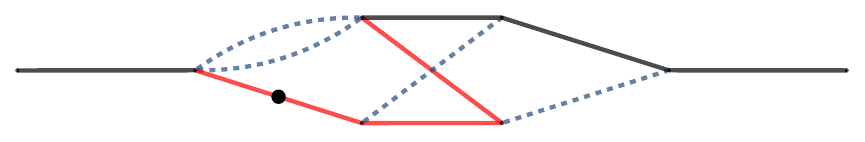}
  \end{tabular}
  \caption{\label{fig::compare_mi}
    Master integrals with nine propagators which appear in the top sector 
    of the differential equations.
    The dashed line denote massless propagators, black and red lines
    have mass $m_b$ and $m_c$, respectively. Squared propagators are
    marked with a dot. The external lines are on-shell. }
\end{figure}

To check the precision of our results obtained with the ``expand and
match'' approach, we compare them with the analytic expressions from
Section~\ref{sub::ana_exp} at $\rho>1/3$.  For the numerical
evaluation of the iterated integrals we use {\tt
  ginac}~\cite{Bauer:2000cp,Vollinga:2004sn}.  As an example, we show
in Fig.~\ref{fig::compare_analytic} the relative difference between
the analytic and numerical result for the $\epsilon^0$ coefficient of
the fourth master integral with nine propagators in
Fig.~\ref{fig::compare_mi}.  The red dots in
Fig.~\ref{fig::compare_analytic} show an agreement of more than 17
digits in the range $0.1<\rho<1$, with increasing precision when
$\rho$ approaches the value $1/2$ where the expansion was matched to the
numerical evaluation with {\tt AMFlow}. Also the expansion around
$\rho=1$ manifests a good convergence when approaching $\rho=1$ because
this series expansion was matched to the analytic boundary condition
calculated in the previous section.  The lower-order $\epsilon$ terms
and the master integrals in lower sectors show in general an even
better agreement.

We observe a loss of precision (only 8 digits) when we approach the
massless limit at $\rho=0$ where several master integrals develop mass
singularities.  In principle the precision can be improved by
including more (regular) expansion points between $\rho=0$ and
$\rho=1/3$.  Alternatively, we can recalculate the boundary condition
with {\tt AMFlow} at some point close to $\rho=0$.  For instance, the
blue points shown in Fig.~\ref{fig::compare_mi} correspond to the
expansion around $\rho=0$ matched to a numerical evaluation at
$\rho=1/100$. In this case, the precision is of more than 44 digits
and thus shows that also in the massless case we can reach a good
accuracy in the evaluation of the master integrals.  

In principle it is possible to use {\tt AMFlow} for the physical value 
of the charm quark mass. However, in that case we would lose the flexibility to
vary the charm quark mass and to consider different mass schemes for the quarks.  
Furthermore, {\tt AMFlow} cannot reproduce the power-log
behaviour around the singular points. In fact, the evaluation with {\tt AMFlow}
at the threshold $\rho=1/3$ or for $\rho=0,1$ yields only the hard parts
in the asymptotic expansion around these two points.  In our approach we
obtain the same power-log expansions around the singular points
as from an analytic calculation, where the expansion coefficients 
have a numerical accuracy of 10 and more.



\section{\label{sec::decay}Decay rates}

We obtain the NNLO prediction to the decay rate after renormalizing the
wave function of the external bottom quark in the on-shell scheme, 
the strong coupling constant $\alpha_s$ in the $\overline{\rm MS}$ scheme with five
active flavour, and $m_c$ and $m_b$ in the on-shell scheme.  
The on-shell masses can be converted to other short-distance mass schemes.
We divide the coefficient $X_2$ appearing in the decay rate at order 
$\alpha_s^2$ in two parts:
\begin{equation}
  X_2(\rho) = X_2^{1c} (\rho) + X_2^{3c}(\rho),
\end{equation}
which correspond to the contributions with one and three charm quarks in the final state.


\subsection{\label{sub::ana_exp}Analytic result of contribution without cut through three charm lines}

We use the results for the master integrals obtained in
Section~\ref{sub::ana_calc} to compute the decay rate at NNLO omitting the
contributions involving three massive charm quarks in the final state,
$X_2^{1c}$. For the
physical values of the charm and bottom quark masses these contributions are
negligible. However, they become sizable in case we approach the limit
$m_c\to 0$ (see next subsection).

The exact expression for $X_2^{1c}$ is too long to be printed here but can be downloaded
from~\cite{progdata}.  It contains in total 313
different iterated integrals constructed from the alphabet in
Eq.~(\ref{eq::alphabet}) up to weight five.  It can be numerically evaluated
using {\tt ginac}~\cite{Bauer:2000cp,Vollinga:2004sn}.

With the help of {\tt HarmonicSums}~\cite{HarmonicSums} we obtain analytic 
expansions around $\rho=0$ and $\rho=1$.  For illustration we
provide the first three expansion terms in both limits. Deeper expansion can
be obtained from~\cite{progdata}.  For the renormalization scale we choose 
$\mu=m_b$ and obtain
\begin{align}
  X_2^{1c} \big|_{\rho \rightarrow 0} =& C_F \bigg\{ \frac{25775}{5184}
  -\frac{13339 \pi ^2}{2592}
  -\frac{101 \zeta_3}{72}
  +\frac{17 }{3} \pi ^2 \log 2
  +\frac{17 \pi ^4}{120}
  +\bigg(\frac{ 13}{8}-\frac{ \pi ^2}{4}+ \zeta_3\bigg) l_\rho 
    \nonumber\\&
  -\frac{5 \pi ^2 }{3}\rho
  +\rho^2 \bigg[
    -\frac{45323}{162}
    +\frac{403 \pi ^2}{54}
    +\frac{599 \zeta_3}{3} 
    -\frac{20}{3} \pi ^2 \log 2
    +\frac{991 \pi ^4}{540}
  \nonumber\\&
    -\left(\frac{290}{9} -\frac{4\pi^2}{3} \right) l_\rho^2
    -\frac{14}{3} l_\rho^3
    -\frac{2}{3} l_\rho^4
    +\bigg(-\frac{  6631}{54}+\frac{52 \pi ^2}{9}+60 \zeta_3\bigg) l_\rho 
    \bigg]  \bigg\}
  \nonumber\\&
  + C_A \bigg\{
  \frac{75623}{5184}
  -\frac{101 \pi ^2}{5184}
  -\frac{1111 \zeta_3}{144}
  -\frac{17}{6} \pi ^2  \log 2
  +\frac{11 \pi ^4}{240}
  -\bigg(\frac{13}{16}-\frac{ \pi ^2}{8}+\frac{ \zeta_3}{2}\bigg) l_\rho
  \nonumber\\&
  +\frac{5 \pi ^2 }{6}\rho
  +\rho^2 \bigg[
    -\frac{56207}{648}
    +\frac{7}{3} l_\rho^3
    +\frac{1}{3}l_\rho^4
    -\frac{745 \pi ^2}{108}
    -\frac{599 \zeta_3}{6}
    +\frac{10}{3} \pi ^2 \log 2
    -\frac{331 \pi ^4}{1080}
    \nonumber\\&
    -\bigg(\frac{   5699}{108}+\frac{26 \pi ^2}{9}+30 \zeta_3\bigg) l_\rho 
    +\bigg(\frac{181}{9}-\frac{2 \pi ^2}{3}\bigg) l_\rho^2 
    \bigg] \bigg\}
  \nonumber\\&
  +  T_F n_l \bigg[ 
    -\frac{1009}{288}
    +\frac{77 \pi ^2}{216}
    +\frac{8 \zeta_3}{3} 
    +\rho^2\bigg(
    \frac{118}{3}-\frac{4\pi ^2}{3}
    +\frac{52}{3}l_\rho-8 l_\rho^2\bigg) 
  \bigg]
  \nonumber\\&
  +  T_F n_b \bigg[
    \frac{16987}{576}
    -\frac{85 \pi ^2}{216}
    -\frac{64 \zeta_3}{3}
    + \rho^2\bigg(-\frac{ 1198}{45}+\frac{8\pi ^2}{3} \bigg)
    \bigg]
  \nonumber\\&
  +  T_F n_c \bigg[
    \frac{20063}{5184}
    +\frac{61 \pi ^2}{216}
    +\frac{4 \zeta_3}{3} 
    +\frac{2}{9} l_\rho^3
    +\frac{5}{3} l_\rho^2
    +\bigg(\frac{415}{72}-\frac{ \pi ^2}{9}\bigg) l_\rho
    -\frac{13 \pi ^2}{8} \rho
    \nonumber\\&
    +\rho^2\bigg(
    -\frac{1475}{162} +\frac{106 \pi ^2}{27} -\frac{184 }{9}l_\rho-\frac{44}{3} l_\rho^2\bigg) 
  \bigg] + \ldots
  \,, 
\label{eq::1c_rho0}
\\
%
  X_2^{1c} \big|_{\rho \rightarrow 1} =& C_F \bigg[ 
    \delta^5 \bigg(
    -\frac{46}{5}+\frac{32 \pi ^2}{5}-\frac{32}{5} \pi ^2 \log 2
    +\frac{48 \zeta_3}{5}\bigg)
    +\delta^6 \bigg(
    \frac{69}{5}-\frac{48 \pi^2}{5}+\frac{48 }{5} \pi ^2 \log 2
    \notag \\&
    -\frac{72 \zeta_3}{5}\bigg)
    +\delta^7 \bigg(
    \frac{39329}{3675}
    +\frac{3044 \pi^2}{945}
    -\frac{496}{105} \pi ^2 \log 2 
    +\frac{248 \zeta_3}{35}
    -\frac{352 }{105}l_{2\delta}
    \bigg)
     \bigg]
  \nonumber\\&
  +  C_A \bigg\{
    \delta^5 \bigg(
    -\frac{286}{15}-\frac{8 \pi ^2}{5}
    -\frac{24 \zeta_3}{5}
    +\frac{16 }{5} \pi ^2 \log 2
    \bigg)
    +\delta^6 \bigg(
    \frac{99}{5}
    +\frac{12 \pi ^2}{5}
    +\frac{36 \zeta_3}{5}
    \nonumber\\&
    -\frac{24 }{5}\pi^2 \log 2 
    \bigg)
    + \delta^7 \bigg[
    -\frac{99547507}{1157625}
    +\frac{62206 \pi ^2}{33075}
    +\frac{132 \zeta_3}{35}
    +\frac{248}{105} \pi ^2 \log 2
    \nonumber\\    &
    + \bigg(\frac{ 1333376}{33075}
    -\frac{256 \pi ^2}{315}
     \bigg)l_{2\delta} 
    -\frac{1408}{315} l_{2\delta}^2
    \bigg]
    \bigg\}
  +  T_F n_l \bigg[ 
    \frac{56 }{15}\delta^5
    -\frac{12 }{5}\delta^6
    \nonumber\\  &
    +\delta^7 \bigg(
    \frac{25577548}{1157625}
    -\frac{512 \pi ^2}{945}
    -\frac{417664}{33075} l_{2\delta}
    +\frac{512 }{315}l_{2\delta}^2
   \bigg)
    \bigg]
  + T_Fn_b \bigg[
    \delta^5 \bigg(\frac{184}{3}-\frac{32 \pi^2}{5} \bigg) 
  \nonumber\\  &
    + \delta^6\bigg(-12+\frac{8 \pi ^2}{5}\bigg)
    +\delta^7 \bigg(\frac{107444}{2835}-\frac{3848 \pi^2}{945} \bigg)
    \bigg]
  +  T_F n_c \bigg[
    \delta^5 \bigg( \frac{184}{3} - \frac{32 \pi^2}{5}\bigg)
  \nonumber\\  &
    +\delta^6 \bigg(-\frac{828}{5} 
    +\frac{88 \pi^2}{5}\bigg) 
    + \delta^7 \bigg( \frac{108580}{567} -\frac{18968 \pi^2}{945}\bigg) 
    \bigg] + \ldots\,,
  \label{eq::1c_rho1}
\end{align}
with $l_\rho = \log(\rho)$ and $l_{2\delta} = \log(2\delta)$. 
The ellipses stand for higher order terms in
$\rho$ and $\delta$.  For the expansion around $\rho=1$, we obtain a
  power-log series starting with $\delta^5$. This is in agreement with
  the results from~\cite{Dowling:2008mc,Fael:2020tow,Fael:2022frj}.
  In the expansion around $\rho=0$ we observe that for the color
  factors $C_FT_Fn_l$ and $C_FT_Fn_b$ the limit $\rho \rightarrow
  0 $ exists whereas the other three colour structures develop
  logarithmic divergences. The non-fermionic structures ($C_F^2$ and
  $C_FC_A$) contain linear logarithms in the leading expansion
  term. For the colour structure $C_FT_Fn_c$ we have a cubic
  logarithm. These logarithms originate from the mass singularities. They cancel
  against the real radiation contribution contained in the three-charm
  contribution $X_2^{3c}$.  At higher order in $\rho$ also quartic
  logarithms start to appear; see, e.g., the colour factors $C_F^2$ and
  $C_FC_A$.
  We note that the coefficients of the odd expansion terms in $\rho$
  are simpler than those of the even terms.

The divergent behaviour of $X_2^{1c}$ for $\rho \to 0$ is due to the mass
singularities for massless charm quarks which are present since not all
possible cuts are considered.  In the complete result as computed in
Refs.~\cite{Pak:2008qt,Pak:2008cp} the limit $\rho\to0$ exists. We can
subtract the expansion in Eq.~(\ref{eq::1c_rho0}) from the result computed in
Refs.~\cite{Pak:2008qt,Pak:2008cp} to obtain analytic expressions for the
contribution from three-charm cuts.  We obtain
\begin{align}
  X_2^{3c} \big|_{\rho \rightarrow 0} =& C_F \bigg\{
    -\frac{409}{576}
    -\frac{349 \pi ^2}{288}
    -\frac{115 \zeta_3}{24} 
    +\frac{19}{6} \pi ^2 \log 2
    -\frac{7 \pi ^4}{144}
    - \bigg(
    \frac{13}{8}-\frac{\pi ^2}{4}+ \zeta_3\bigg)l_\rho 
    +\frac{5 \pi ^2 }{3}\rho
    \nonumber\\&
    +\rho^2 \bigg[
    \frac{12083}{648}
    -\frac{103 \pi ^2}{36}
    -\frac{341 \zeta_3}{3}
    -\frac{4}{3} \pi ^2 \log 2
    -\frac{29 \pi ^4}{18}
    + \bigg(
      \frac{961}{54}-\frac{52 \pi ^2}{9}-60\zeta_3
    \bigg)l_\rho 
    \nonumber\\&
    -\bigg(\frac{34}{9}+\frac{4\pi^2}{3}\bigg)l_\rho^2 
    +\frac{14 }{3}l_\rho^3
    +\frac{2 }{3}l_\rho^4
    \bigg]
    \bigg\} 
  + C_A \bigg\{
    \frac{409}{1152}
    +\frac{349 \pi ^2}{576}
    -\frac{19}{12} \pi ^2 \log 2
    \nonumber\\&
    +\frac{115 \zeta_3}{48}
    +\frac{7 \pi ^4}{288}
    -\frac{5 \pi ^2 }{6}\rho 
    +\bigg(
    \frac{13}{16}
    -\frac{ \pi ^2}{8}
    +\frac{ \zeta_3}{2}\bigg)l_\rho  
    +\rho^2 \bigg[
    -\frac{12083}{1296}
    +\frac{103 \pi ^2}{72}
    +\frac{341 \zeta_3}{6} 
    \nonumber\\&
    +\frac{2}{3} \pi^2 \log 2
    +\frac{29 \pi ^4}{36}
    - \bigg(
    \frac{961}{108}-\frac{26 \pi ^2}{9}-30 \zeta_3
    \bigg)l_\rho 
    +  \bigg(
    \frac{17}{9}
    +\frac{2 \pi ^2}{3}\bigg)l_\rho^2
    -\frac{7 }{3}l_\rho^3
    -\frac{1}{3}l_\rho^4
    \bigg]
    \bigg\}
    \nonumber\\&
  + T_F n_c \bigg[
    -\frac{38225}{5184}
    +\frac{2 \pi ^2}{27}
    +\frac{4 \zeta_3}{3}
    +\bigg(
    -\frac{415}{72}+\frac{ \pi ^2}{9}\bigg)l_\rho 
    -\frac{5 }{3}l_\rho^2
    -\frac{2 }{9}l_\rho^3
    +\frac{3 \pi ^2 }{8}\rho 
    \nonumber\\&
    + \rho^2\bigg(
    \frac{9305}{162}
    +\frac{38 \pi ^2}{27}
    +\frac{340 }{9}l_\rho
    +\frac{20 }{3}l_\rho^2
    \bigg)
    \bigg] +\ldots \,.
  \label{eqn:X2^3c}
\end{align}
A deeper expansion is again available from~\cite{progdata}.
After specifying the QCD colour factors to QED, the result in Eq.~\eqref{eqn:X2^3c}
yields the LO branching ratio also for the rare muon decay
$\mu \to e(e^+e^-) \nu_\mu \bar \nu_e$ in an analytic form.
We find perfect agreement with the numerical results for
the branching ratio in Refs.~\cite{Fael:2016yle,Pruna:2016spf}.


\subsection{Numeric results of complete contributions}

The ``expand and match'' approach described in Section~\ref{sub::numeric}
provides results for the whole $\rho$ range as (power-log) expansions around
values $\rho_i$ with numerical coefficients. Each expansion is used in the
respective convergence region.

There are a number of checks of our results which we describe in the
following. A strong check of the individual master integrals is
provided by the comparison to results obtained with {\tt AMFlow} at
various values for $\rho$. The numerical results we obtain from {\tt
  AMFlow} are compared to the values we obtain from the power-log
expansion of the masters.  Looking at the relative difference, we
observe that in the region between $\rho=1$ and the threshold at
$\rho=1/3$ the accuracy for the Taylor expansions is of the order of
$10^{-35}$ to $ 10^{-40}$. When we match to the expansion around
threshold, we loose some precision (because of the more complicated
expansion) and observe for the Taylor expansions a relative difference
of the order of $10^{-15}$ to $10^{-20}$ in the region between the
threshold and $\rho=0$.  After matching to the power-log expansion
around $\rho=0$ we loose again some precision. Nevertheless we can
reproduce the coefficients of the analytic expansion from
Refs.~\cite{Pak:2008qt,Pak:2008cp} to 8 or more digits, see also
below.

The bare NNLO expression contains spurious $\epsilon$ poles which are
introduced during the integration-by-parts reduction.  After inserting
results for the master integrals we observe poles up to sixth
order. Since the counterterm contributions provide at most
$1/\epsilon^2$ poles the higher poles have to cancel in the bare NNLO
result.  In the worst case we observe a pole cancellation of order
$10^{-15}$ for certain $1/\epsilon^3$ poles while for the higher poles
the cancellation is much better.

\begin{figure}[t]
    \centering
    \begin{tabular}{c}
      \includegraphics[width=.9\textwidth]{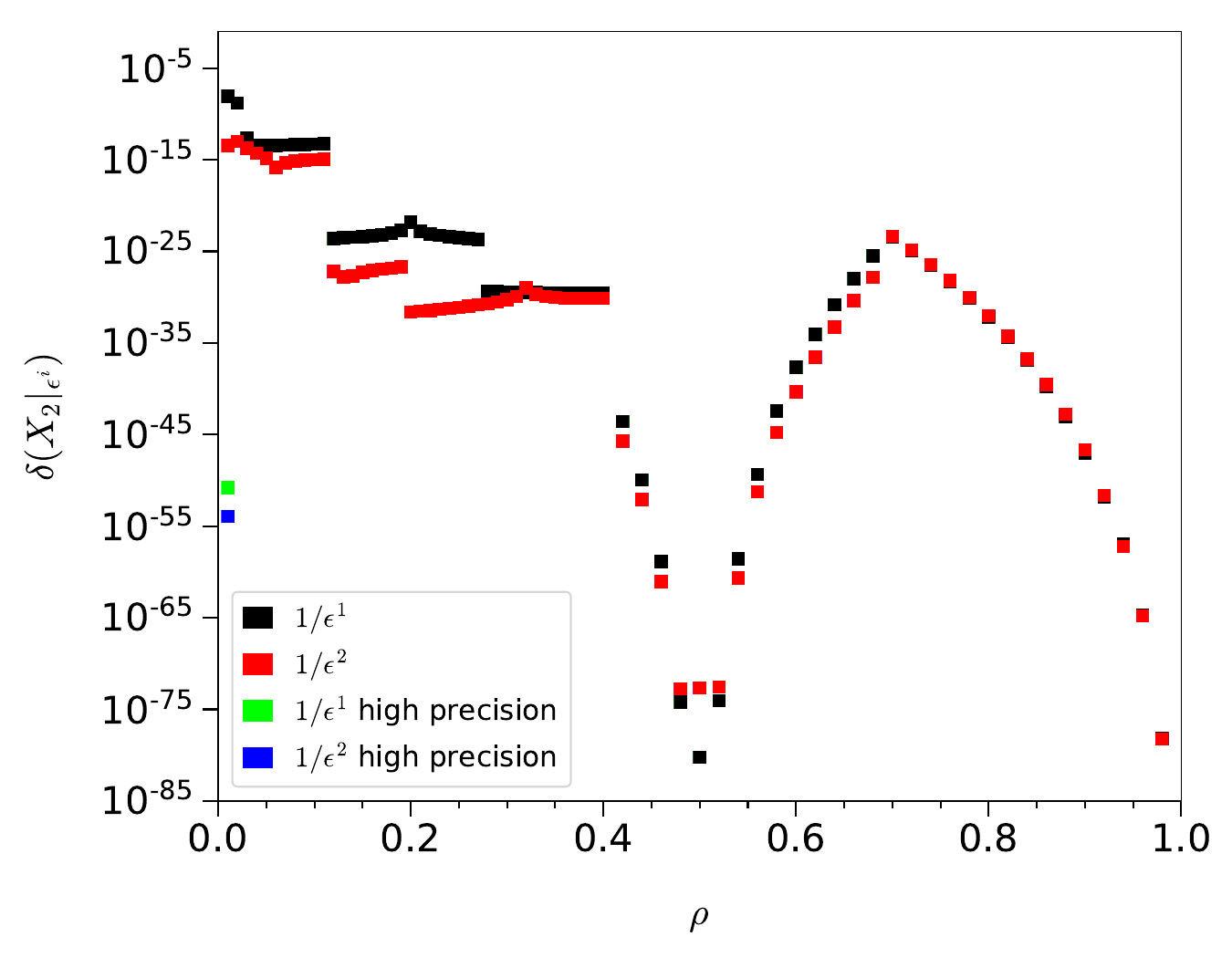}
    \end{tabular}
    \caption{\label{fig::pole_cancellation} Relative precision for the
      cancellation of the $1/\epsilon$ poles according to
      Eq.~(\ref{eq::pole_cancellation}).  All colour factors have been
    set to their numerical values.}
  \end{figure}

We obtain the renormalized decay rate by adding the counterterm in
analytic form. The cancellation of the poles in $\epsilon$ provides a check on
the numerical precision of our result. To quantify the accuracy we introduce
the quantity
\begin{equation}
  \delta (X_2\big|_{\epsilon^i}) = \left| \frac{X_2^{\rm
      bare}\big|_{\epsilon^i} + X_2^{\rm CT}\big|_{\epsilon^i}}
         {X_2^{\rm CT}\big|_{\epsilon^i}} \right|\,.
         \label{eq::pole_cancellation}
\end{equation}
In Fig.~\ref{fig::pole_cancellation} we show $\delta
(X_2\big|_{\epsilon^{-2}})$ and $\delta (X_2\big|_{\epsilon^{-1}})$
for $0\le \rho \le 1$. Close to the starting point at $\rho=1/2$ both
the linear and quadratic poles cancel with more than 70 digits and
then accuracy deteriorates when the distance from $\rho=1/2$
increases. Here the accuracy is limited by the truncation of the
expansion around $\rho=1/2$.

When moving towards $\rho=0$ the accuracy slightly deteriorates.
Close to $\rho=0$ we observe still a precision of about
8 digits.  The blue and green points in
Fig.~\ref{fig::pole_cancellation} shows the precision in the
$1/\epsilon^2$ and $1/\epsilon$ poles in case we use the $\rho=0$
expansion matched to the {\tt AMFlow} evaluation at $\rho=1/100$.

\begin{figure}[t]
  \centering
  \begin{tabular}{c}
    \includegraphics[width=.9\textwidth]{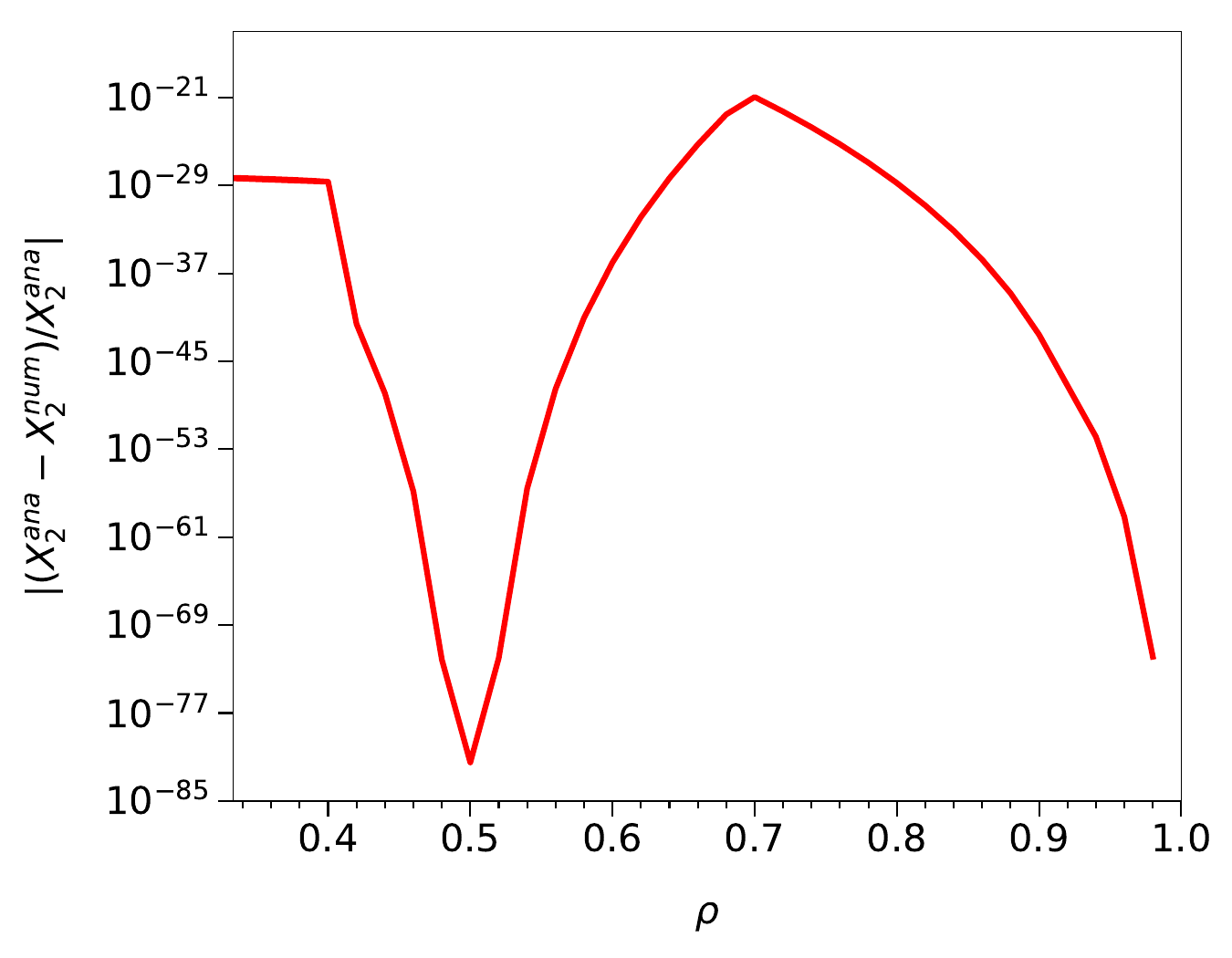}
  \end{tabular}
  \caption{\label{fig::compare_amp_analytic}Relative difference between the
    semi-analytic approximation and the analytic results for the NNLO
    contribution $X_2$ to the decay rate.  For the comparison we have set
    all colour factors to their numerical values. We consider the region
    $1/3<\rho<1$ since for $\rho<1/3$ there are three-charm contributions
    which are not contained in the analytic result.}
\end{figure}

  To check the numerical accuracy of our approximation we compare for $\rho>1/3$
  against analytic results presented in the previous subsection. 
  In Fig.~\ref{fig::compare_amp_analytic} we show for the decay rate the
  relative difference between the analytic result
  and the semi-analytic approximation. For the ``expand and match'' approach
  we have used boundary conditions obtained with {\tt AMFlow} for
  $\rho=1/2$. Over the whole range of $\rho$ we observe an agreement of
  at least 19 digits.  

A further strong check of our result comes from the comparison of our
expansion for $\rho\to 0$ to the results of
Refs.~\cite{Pak:2008qt,Pak:2008cp}. Using boundary conditions from $\rho=1/2$
and transporting the solution of the master integrals at $\rho=0$, we can
reproduce the analytic coefficients of the asymptotic expansion at $\rho=0$
with 10 digits for $n=0$ which decreases to 8 for $n=5$.

Alternatively we can also use the $\rho=0$ expansion matched at
$\rho=1/100$. In this case we reproduce the analytic
coefficients~\cite{Pak:2008qt,Pak:2008cp} with more than 50 digits.  Note that
in our approach it is possible to obtain without any effort 50 expansion terms
for $\rho\to0$.

\begin{figure}[t]
  \centering
  \begin{tabular}{cc}
    \includegraphics[width=.45\textwidth]{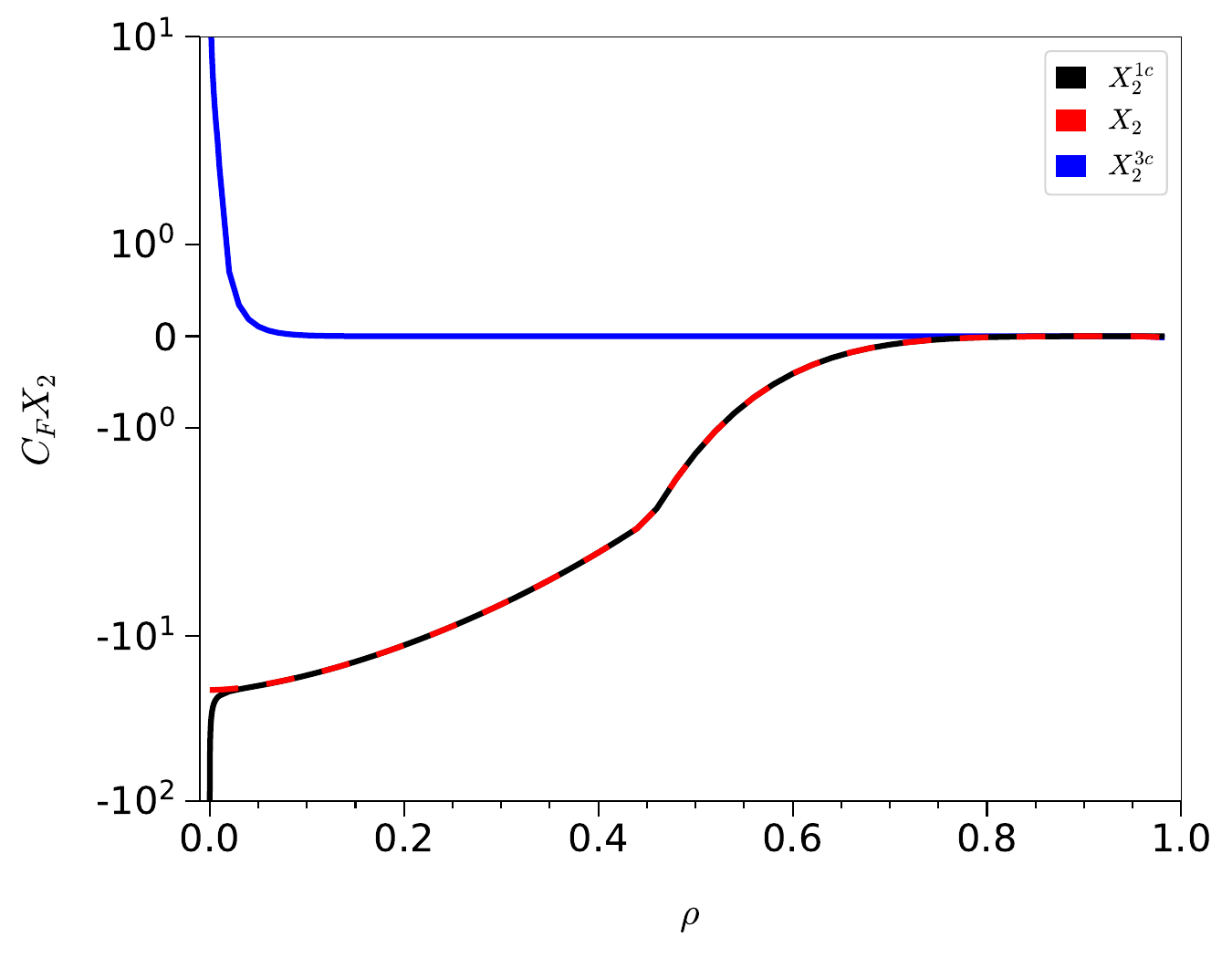} &
    \includegraphics[width=.45\textwidth]{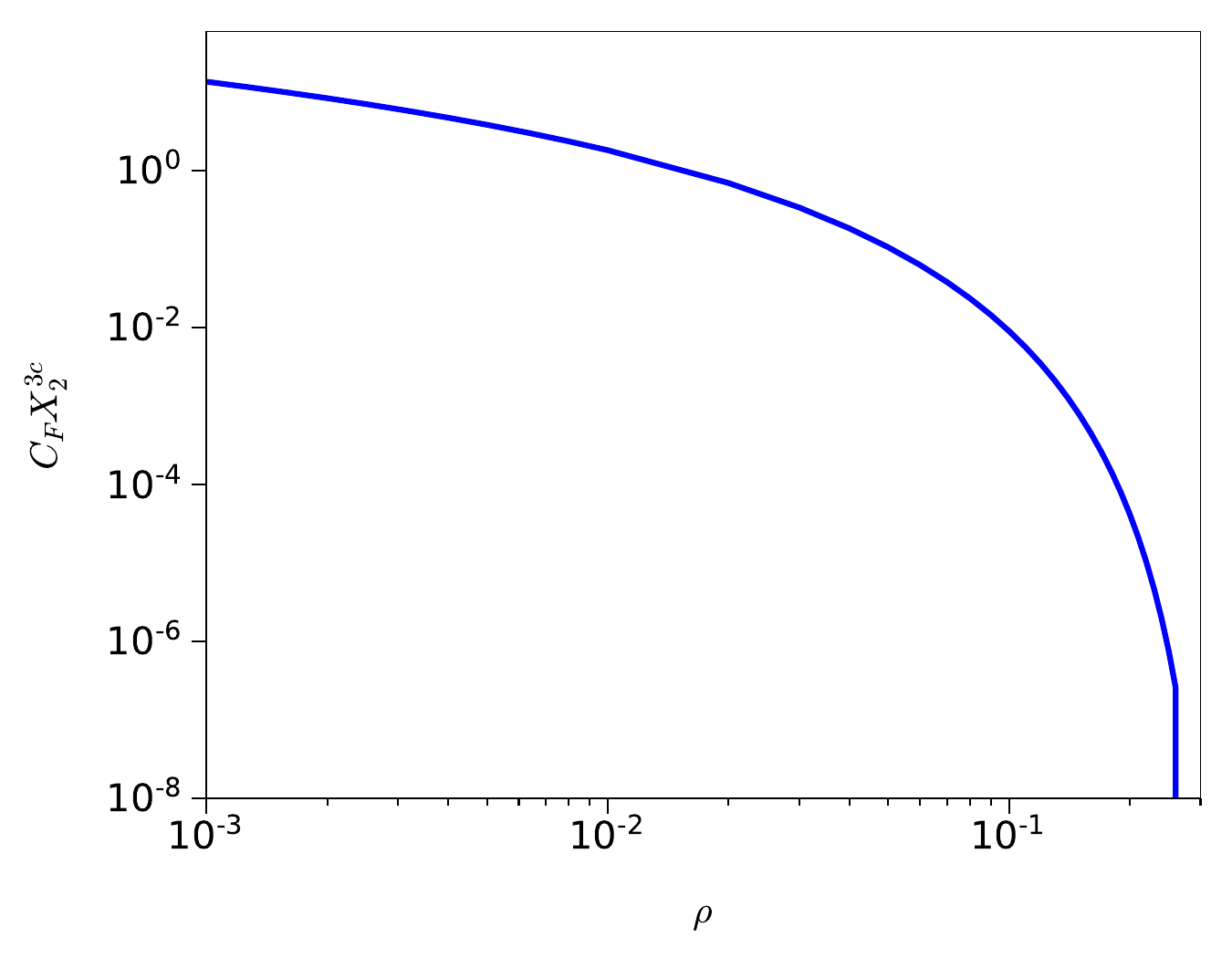}
  \end{tabular}
  \caption{\label{fig::NNLO}The complete result (dashed, red) 
    is compared with the individual contributions which only contain
    cuts through one (black) or three charm quark lines (blue), respectively.
    The right panel shows the three-charm contribution is a double
    logarithmic plot.
  }
\end{figure}

In Fig.~\ref{fig::NNLO} we show the NNLO contribution $X_2$ to the
decay rate. The complete result is shown in red. For comparison we
also show the contribution which only contains cuts through one charm
quark (black curve, see Section~\ref{sub::ana_calc}). The strong
$\log^3\rho$ behaviour for small values of $\rho$ is clearly visible,
however, only very close to $\rho=0$. The blue curve represents the
contribution with three charm quarks in the final state. It has
the same logarithmic behaviour with an opposite sign
such that after adding it to the black
curve one obtains the complete NNLO corrections with a smooth limit for
$\rho\to 0$.

Finally we note that the three charm contribution  to the decay rate 
is extremely suppressed at the physical value of the charm and bottom masses
and therefore irrelevant for the current accuracy in the extraction of $|V_{cb}|$.
At $\rho=0.2$ we have $X_2^{3c} = 4 \times 10^{-5}$ which yields 
$\mathrm{Br}(b \to c \bar c c \ell \bar \nu_\ell) = 4 \times 10^{-8}$.

\section{\boldmath Charm quark contribution in $b\to u$ decay rate at NNLO}
\label{sec:Uc}

In this Section we compute the charm quark mass dependence to
$b\to u \ell \bar{\nu}_\ell$ at NNLO which arises from diagrams as the one
shown in Fig.~\ref{fig::1c3c}(i).  
In analogy to Eq.~(\ref{eq::gamb2c}) we write
\begin{eqnarray}
  \Gamma(B \to X_u \ell \bar\nu) 
  &=& \Gamma_0 \left[1 + {\left(\frac{\alpha_s}{\pi}\right)^2 C_F T_F} X_2^{C} + \ldots
      \right] 
      + {\cal O}\left(\frac{\Lambda_{\rm QCD}^2}{m_b^2}\right)\,,
      \label{eq::gamb2u_charm}
\end{eqnarray}
with $T_F=1/2$. The ellipses stand for charm quark-independent
contributions.  In the following we discuss the results for $X_2^{C}$.

In total, there are four Feynman
diagrams. After integration-by-parts reduction, we find 16 master integrals.
For the computation of the master integrals we again apply the ``expand and
match'' approach and use {\tt AMFlow} in order to obtain the boundary
conditions.  In contrast to $b\to c \ell \bar{\nu}_\ell$
we have cuts through two charm quarks and thus the threshold is located at
$\rho=1/2$ instead of $\rho=1/3$.  This means the singular points are
$\rho=0$ and $1/2$.
For the expansion around $\rho_0=0$ we can use
the ansatz given in Eqs.~(\ref{eqLLAnsatzdel0}). For $\rho_0=1/2$ a new ansatz is necessary since we expect the
occurrence of square roots according to Eq.~(\ref{eq::v_pow}). We therefore
have
\begin{align}
  I_i = \sum_{j=\epsilon_{\rm min}}^{\epsilon_{\rm max}}
  \sum_{m=0}^{j+4} \sum_{n=n_{\rm min}}^{n_{\rm max}} 
  c_{i,j,m,n} \, \epsilon^j \left(\sqrt{\rho - \rho_0}\right)^n \log ^m
  \left(\sqrt{\rho-\rho_0} \right) , 
\end{align}
where $\rho_0=1/2$.  Again we allow for negative values of $n_{\rm
  min}$. However, the solution for the differential equations requires
$n\ge0$.  We note that the additional imaginary parts induced by the
two particle threshold are now generated by both the square roots and
the logarithms.

\begin{figure}[t]
  \centering
  \begin{tabular}{c}
    \includegraphics[width=.9\textwidth]{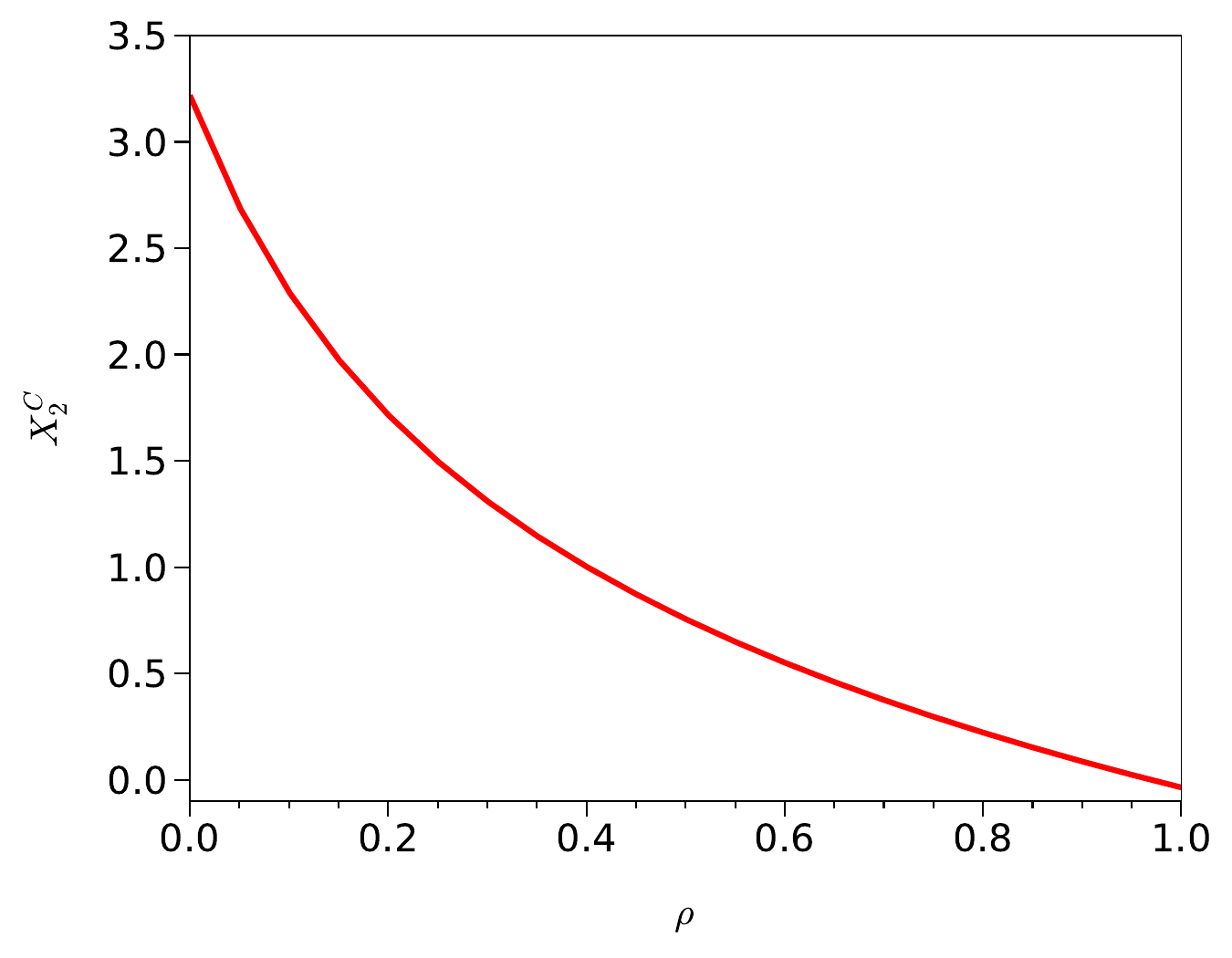}
  \end{tabular}
  \caption{\label{fig::UC}$X_2^{C}$ as a function of $\rho$.}
\end{figure}

Finite results for the decay rate are obtained after renormalizing the
bottom wave function in the on-shell and $\alpha_s$ in the
$\overline{\rm MS}$ scheme. Results for this contribution are given in
Eq.~(A6) of Ref.~\cite{Pak:2008cp} in an expansion for $\rho\to 0$ up
to $\rho^7$. We use numerical boundary conditions at
$\delta^\prime=0.1$ ($\rho \approx 0.95$) and apply repeatedly
the ``expand and match'' approach to go over the two-charm threshold
and to obtain a semi-analytic expansion around $\rho=0$ (including 50
expansion terms).  We can compare the individual coefficients of
$\rho^n \log^k(\rho)$ to the analytic results of
Ref.~\cite{Pak:2008cp} and find agreement to at least 9 significant
digits.  For $\rho=1$ we can compare to the $n_b$ term of the
$b\to c$ decay which is obtained from Eq.~(\ref{eq::1c_rho0}) after
setting $\rho=0$.  We observe agreement at the level of $10^{-30}$.
Similarly, for $\rho=0$ we compare to the $n_l$ term of
Eq.~(\ref{eq::1c_rho0}) and find agreement within 9 digits.
This shows again the power of our approach. Let us again stress that
our numerical precision can be systematically improved by choosing
more expansion points $\rho_0$, deeper expansion depths or an
appropriately chosen value for $\rho$ to compute the boundary terms.
In Fig.~\ref{fig::UC} we show $X_2^{C}$ as a function of $\rho$.

\section{\label{sec::concl}Conclusions and outlook}

In this work we reconsider the NNLO corrections to the semileptonic decay of a
bottom quark to a charm quark. In the literature expansions are
available~\cite{Pak:2008qt,Pak:2008cp,Dowling:2008mc} which are sufficient for
phenomenological studies.  The aim of this paper is to provide semi-analytic
power-log expansions which are valid for $0 \le \rho=m_c/m_b \le 1$.
Furthermore, we provide analytic results for the subset of Feynman diagrams
which have contributions with one charm quark in the final state.

For the semi-analytic calculation we use the ``expand and
match''~\cite{Fael:2021kyg,Fael:2022miw} method to transport information about
the master integrals at a given starting point $\rho=\rho_0$ to any value of
$\rho$. It uses the differential equations in combination with an appropriate
ansatz to construct a semi-analytic approximation formula which is composed of
power-log expansions valid is certain sub-intervals of $\rho\in[0,1]$.  To
obtain the initial values of the master integrals we use {\tt
  AMFlow}~\cite{Liu:2022chg}. Our approach is able to properly take into
account singular behaviours of the exact (unknown) function. In our case this
concerns the expansions around $\rho=0, 1/3$ and $1$.  We also compute the
charm dependent contributions to $b \to u \ell \bar \nu_\ell$ at NNLO. Here
the singular points are at $\rho=0$ and $1/2$.

The method developed in this paper serves as preparation for the computation
of non-leptonic decay rates at NNLO.  In these cases the techniques used for
the semileptonic decays to obtain expansions are either not applicable or
technically quite challenging. On the other hand it is straightforward to
extend the semi-analytic approach of the present paper.

We want to remark that the method described in this paper can be applied
at N$^3$LO if the reduction to master integrals is possible and the system of
differential equations can be established.  However, it seems that at the
moment the latter is a serious bottleneck.



\section*{Acknowledgements}  

We thank F.\ Lange for the help with {\tt Kira}, V.\ Shtabovenko for 
the help with {\tt Canonica} and K.\ Melnikov for useful discussions.
This research was supported by the Deutsche Forschungsgemeinschaft (DFG,
German Research Foundation) under grant 396021762 --- TRR 257 ``Particle
Physics Phenomenology after the Higgs Discovery'' and has received funding
from the European Research Council (ERC) under the European Union's Horizon
2020 research and innovation programme grant agreement 101019620 (ERC Advanced
Grant TOPUP).  
The work of M.F. is supported by the European Union’s Horizon 2020 
research and innovation program under the Marie Sk\l{}odowska-Curie grant agreement 
No.\ 101065445 -- PHOBIDE.


\end{document}